\documentclass{article}
\usepackage{jheppub}

\usepackage{ifpdf}

\usepackage{afterpage}

\usepackage{amsmath,amssymb}
\usepackage{graphicx}
\usepackage{multirow}
\usepackage{arydshln}
\usepackage{bbm}
\usepackage[normalem]{ulem}
\usepackage{subfigure}
\usepackage{arydshln}
\usepackage{feynmp}
\usepackage{color, colortbl}
\definecolor{LightCyan}{rgb}{0.88,1,1}
\definecolor{LightViolet}{rgb}{1, 0.88,1}
\definecolor{LightGreen}{rgb}{0.88, 1, 0.88}

\let\tmptitle\title\renewcommand{\title}[1]{\tmptitle{\Large #1}}
\let\tmpdate\date\renewcommand{\date}[1]{\tmpdate{\normalsize #1}}

\newcommand{\nn}{\nonumber}
\allowdisplaybreaks[2]
\newcommand{\ba}{\begin{eqnarray}}
\newcommand{\ea}{\end{eqnarray}}
\newcommand{\be}{\begin{equation}}
\newcommand{\ee}{\end{equation}}

\newcommand{\bear}{\begin{array}}
\newcommand{\ear}{\end{array}}

\newcommand{\beq}{\begin{eqnarray}}
\newcommand{\eeq}{\end{eqnarray}}
\newcommand{\beqa}{\begin{eqnarray}}
\newcommand{\eeqa}{\end{eqnarray}}

\def\OMIT#1{{}}
\newcommand{\lsim}{\mathrel{\rlap{\lower4pt\hbox{\hskip1pt$\sim$}}
    \raise1pt\hbox{$<$}}}         
\newcommand{\gsim}{\mathrel{\rlap{\lower4pt\hbox{\hskip1pt$\sim$}}
    \raise1pt\hbox{$>$}}}         

\def\lsim{\mathrel{\rlap{\lower4pt\hbox{\hskip1pt$\sim$}}
    \raise1pt\hbox{$<$}}}         
\def\gsim{\mathrel{\rlap{\lower4pt\hbox{\hskip1pt$\sim$}}
    \raise1pt\hbox{$>$}}}         

\begin{document}

\vspace*{-30mm}

\title{Fitting the Higgs to Natural SUSY}

\author[a,b]{Raffaele Tito D'Agnolo,}
\author[c]{Eric Kuflik,}
\author[d]{Marco Zanetti}
\affiliation[a]{CERN, European Organization for Nuclear Research, Geneva, Switzerland}
\affiliation[b]{Scuola Normale Superiore and INFN Pisa, Piazza dei Cavalieri 7, 56126, Pisa, Italy}
\affiliation[c]{Raymond and Beverly Sackler School of Physics and Astronomy, Tel-Aviv University, Tel-Aviv 69978, Israel}
\affiliation[d]{Massachussets Institute of Technology, Cambridge, MA, USA}
\emailAdd{raffaele.dagnolo@sns.it}
\emailAdd{ekuflik@post.tau.ac.il}
\emailAdd{marco.zanetti@cern.ch}

\vspace*{1cm}

\abstract{
We present a fit to the 2012 LHC Higgs data in different supersymmetric frameworks using naturalness as a guiding principle. We consider the MSSM and its $D$-term and $F$-term extensions that can raise the tree-level Higgs mass.  When adding an extra chiral superfield to the MSSM, three parameters are needed determine the tree-level couplings of the lightest Higgs. Two more parameters cover the most relevant loop corrections, that affect the $h\gamma\gamma$ and $hgg$ vertexes. Motivated by this consideration, we present the results of a five parameters fit encompassing a vast class of complete supersymmetric theories. We find meaningful bounds on singlet mixing and on the mass of the pseudoscalar Higgs $m_A$ as a function of $\tan\beta$ in the MSSM.  We show that in the $(m_A, \tan\beta)$ plane, Higgs couplings measurements are probing areas of parameter space currently inaccessible to direct searches. We also consider separately the two cases in which only loop effects or only tree-level effects are sizeable. In the former case we study in detail stops' and charginos' contributions to Higgs couplings, while in the latter we show that the data point to the decoupling limit of the Higgs sector. In a particular realization of the decoupling limit, with an approximate PQ symmetry, we obtain constraints on the heavy scalar Higgs mass in a general type-II Two Higgs Doublet Model.}
\maketitle

\section{Introduction}
With the 2012 run reaching its conclusion our knowledge of Higgs couplings is considerably improving. Some purists might argue that the new resonance~\cite{:2012gu, :2012gk} is not yet experimentally proven to be the Higgs boson, but all evidence goes in that direction and even more so after the recent updates by the LHC experiments~\cite{CMSH, ATLASHComb}. In the previous round of LHC results, namely 4.7 $\mathrm{fb}^{-1}$ at $\sqrt{s}=7$ TeV and 5.9 $\mathrm{fb}^{-1}$ at $\sqrt{s}=8$ TeV, the errors in the Higgs' couplings were dominated by statistical uncertainties that are expected to scale-down with increased luminosity.  Nonetheless, these measurements were already precise enough to make meaningful statements about broad classes of models \cite{Carmi:2012in, Ellis:2012rx, Espinosa:2012ir, Montull:2012ik, Bertolini:2012gu}. After the addition of approximately 7 $\mathrm{fb}^{-1}$ at $\sqrt{s}=8$ TeV, we find it interesting to explore the consequences of Higgs measurements in the predictive and theoretically motivated framework of natural supersymmetry. Several groups have already fitted Higgs' rates in different contexts, focusing mainly on simplified settings in which all the couplings were determined by one or two free parameters~\cite{Espinosa:2012in, Carmi:2012yp, Azatov:2012bz, Espinosa:2012im, Giardino:2012ww}, concentrating also on the cases of light stops or the type-II two Higgs doublet model (2HDM)~\cite{Carmi:2012in, Espinosa:2012in}. Here we adopt a different perspective, discussing the implications of Higgs measurements in complete natural supersymmetric theories.

In Section \ref{sec:predict} we briefly review our definition of natural supersymmetry. In this context, it is necessary to extend the Higgs sector of the MSSM to obtain the observed Higgs mass $m_h\approx 126$~GeV without incurring excessive fine-tuning. Therefore we consider $D$-term and $F$-term models that modify the 2HDM structure of the MSSM potential.

We study the new tree-level effects, distinguishing between the cases in which an approximately type-II 2HDM structure is preserved and those in which the mixing with extra light states in the Higgs sector is sizeable. We find that in general three couplings are enough to parameterize all the tree-level deviations in Higgs rates. Loop corrections to $hgg$ and $h\gamma\gamma$  vertices can also be sizeable, since naturalness considerations require stops and Higgsinos to be fairly light. Two extra parameters are needed to include these effects. Therefore we perform a five dimensional fit to the data. Although naturalness has driven us to consider a certain spectrum and a certain subset of supersymmetric effects on Higgs couplings, the fine-tuning constraints are not imposed in fits. The general five dimensional framework parameterizes a vast class of complete theories, that may or may not be fine-tuned.
We also present fit results in several simplified settings, considering the cases in which: (1) only the tree-level mixing with the second Higgs doublet plays a role, (2) only tree-level effects are important, but both the second Higgs doublet and an extra singlet are responsible for deviations in the couplings and (3) only MSSM loop effects from stops and charginos are sizeable.

The fitting procedure is described in Section~\ref{sec:fit} together with the data taken as input. In Section~\ref{sec:res} we present the results of the fit and discuss their relevance in light of direct searches performed at the LHC.

\section{The Natural SUSY framework}
\label{sec:predict}
Deviations in the Higgs couplings are probed by the measurements of the Higgs rates, which are determined by partial widths.  We denote the modifications to the partial widths as
\be\label{eq:r}
\left| r_{i} \right|^2  \equiv \frac{\Gamma(h\rightarrow i i  )~~~}   {~\Gamma(h\rightarrow i i )^{\rm SM}},
\ee
with $i = t, V, G, \gamma, b, \tau$ standing for top, massive vector gauge boson, gluon, photon, bottom and tau, respectively. In general it is hard to write the $r_i$'s in terms of few supersymmetry breaking parameters. However if we restrict to natural theories it is possible to identify a limited number of relevant effects, as was shown in~\cite{Blum:2012ii}. As stated above, we do not impose fine-tuning bounds in the fits, but we still find naturalness as a good guiding principle to select models and simplified scenarios in which fit results can be interpreted.

For concreteness, consider at most a $\Delta^{-1}=10\%$ tuning, where
\be
\Delta \equiv \left|\frac{2\delta m_{H_u}^2}{m_h^2}\right|\, , \quad
 \delta m_{H_u}^2|_{\mathrm{stops}} = -\frac{3}{8\pi^2}y_t^2\left(m_{\tilde Q_3}^2+m_{\tilde u_3}^2+|A_t|^2\right)\log\frac{\Lambda}{\mathrm{TeV}}\, . \label{eq:ft}
\ee
This restricts, in the MSSM, an unmixed stop to be lighter than about 600 GeV. Taking into account also the relation between the Higgs mass and the Higgsino mass: $-m_h^2/2=|\mu|^2+...$, limits one chargino to be lighter than 300 GeV. More precisely, the upper bounds obtained from fine-tuning take the form~\cite{Papucci:2011wy}
\ba
\mu &\lesssim& 290\;\mathrm{GeV} \frac{m_h}{125\; \mathrm{GeV}}\sqrt{\frac{10\%}{\Delta^{-1}}}\, , \nn \\
\sqrt{m_{\tilde t_1}^2+m_{\tilde t_2}^2}&\lesssim& 880 \; \mathrm{GeV}\frac{\sin \beta}{\left(1+a_t^2\right)^{1/2}} \sqrt{\frac{3}{\log(\Lambda/\mathrm{TeV})}}\frac{m_h}{125\; \mathrm{GeV}}\sqrt{\frac{10\%}{\Delta^{-1}}}\, ,
\ea
where $a_t=A_t/\sqrt{m_{\tilde t_1}^2+m_{\tilde t_1}^2}$ and $\Lambda$ is the scale at which supersymmetry breaking effects are mediated to the MSSM. At tree level, the bound on $\mu$ translates into a bound on the lightest chargino
\be
m_{\chi^\pm_1} \le \sqrt{|\mu|^2 + M_W^2} \lesssim 300 \; {\rm GeV}.
\ee
Other Standard Model (SM) superpartners can be much heavier, with masses ranging from a few TeVs to tens of TeVs.

However, raising the Higgs mass to 126 GeV in the MSSM through stop radiative corrections, implies a tuning much worse than 10\%~\cite{Draper:2011aa,Hall:2011aa}. To accommodate a 126 GeV Higgs boson, natural supersymmetry must be non-minimal, allowing for additional interactions that modify the Higgs sector. The new interactions deform the Higgs quartic potential at tree level and change the Higgs couplings beyond the significant deviations (relative to the SM) that can already occur in the MSSM. In the following we will consider both extensions that leave the type-II 2HDM structure of the potential unaltered and shift quartic couplings relative to the MSSM, and models in which the 2HDM relations between light Higgs couplings are relaxed.

Nonetheless, naturalness not only complicates the problem, as seen above, it  also limits the size of $A_t$ and $\mu$. This greatly simplifies the problem of determining the $r_i$'s as a function of soft supersymmetry breaking parameters. First and foremost, the type-II 2HDM structure of the MSSM Higgs potential is not strongly affected by non-holomorphic loop corrections in a natural theory, regardless of our definition of tuning. As an example we can take the definition at the beginning of this section and ask for a $10\%$ tuning, then the corrections to the bottom Yukawa for $\tan\beta\leq 40$ (and $r_b = \mathcal{O}(1)$ at tree-level) are negligible \cite{Blum:2012ii}. So in the natural MSSM and its $D$-term extensions, Higgs couplings can be expressed in terms of only two parameters at tree-level, as in a general type-II 2HDM. Adding a singlet chiral superfield, as for example in the NMSSM, increases the number of parameters to three, if we assume all couplings to be CP conserving. The additional degree-of-freedom parameterizes the mixing between the new CP even state and the lightest scalar Higgs: $c_\phi \equiv \langle S | h \rangle$.

We have seen that naturalness also prefers stops and charginos to be fairly light, and for this reason we consider their loop contributions to the Higgs to digluon/diphoton partial widths, introducing two new parameters. They can be chosen as $\delta r^{\tilde{t}}_G$, characterizing the stop contribution to the dimension-5 Higgs-gluon-gluon coupling and $\delta r^{\tilde{\chi}^\pm}_\gamma$, characterizing the chargino contribution to the dimension-5 Higgs-photon-photon-coupling.  It was suggested that light staus can be responsible for the enhancement of the $h\to\gamma\gamma$ rate observed at the LHC~\cite{Carena:2011aa, Carena:2012gp, Giudice:2012pf}, but we do not consider this possibility explicitly in what follows. We also neglect sbottom loop corrections and charged Higgs loop corrections. In all three cases we expect contributions larger than a few \% only in extreme corners of the parameter space that imply a tuning much worse than $10\%$ in our definition~\cite{Blum:2012ii}. In this regions charge breaking minima can also be generated~\cite{Giudice:2012pf}. Nonetheless, in the 5 dimensional fits, $r_G$ and $r_\gamma$ are treated as independent parameters. Thus we capture any loop effect on these vertexes.

In terms of the five parameters discussed above, the partial width modifiers are given by
\be\begin{array}{rcl}
r_\tau &=&r_b\\
r_V    &=&(1-c^2_\phi+r_b r_t)/(r_t+r_b) \\
r_G    &=& r_t(1 +\delta r^{\tilde{t}}_G) \\
 r_\gamma &\approx& 1.27r_V-0.27r_G+ \delta r^{\tilde{\chi}^\pm}_\gamma\, ,
\label{eq:coupling}
\end{array}\ee
where we have chosen $r_b$ and $r_t$ to parametrize the 2HDM tree-level effects.
Note that $r_\tau=r_b$ assumes that non-holomorphic corrections to the Higgs interactions with down quarks and leptons are small relative to the tree-level value, as discussed above. Here, and in what follows, we always neglect Higgs decays to non-SM particles, in particular to neutralinos.
In Section~\ref{sec:2hdm} we review all tree-level predictions in more detail, starting with a generic 2HDM with $c_\phi=0$, and later introducing a superpotential for the singlet. In Sections~\ref{sec:stops} and~\ref{sec:charginos} we study the bounds on loop corrections from naturalness and direct searches and compare them with the results of the fit.

\section{Data and fitting procedure}
\label{sec:fit}
In order to constrain the supersymmetric parameters, we take into account all the available Higgs channel rates \cite{CMSH, ATLASHComb}.  The rates depend on the product of the overall production cross-section and branching ratio for the particular channel. The results are typically reported as a confidence interval on the event rate relative to the SM prediction, denoted by $\hat{\mu}$.  We take all the $\hat{\mu}$'s at the two LHC best fit values for the mass: $m_h=125.8$ GeV for CMS and $m_h=126.0$ GeV for ATLAS. For the Tevatron $h\to b\bar b$ rate we take the value of $\hat{\mu}$ at $m_h=126.0$~GeV.
 In Table~\ref{tab:ratesvalues} we list the analyses relevant for our fit, regrouped by channel, and the corresponding values of the $\hat{\mu}$'s.

In any given model, the signal strengths are determined by the $ r_{i}$'s. For the LHC, the four relevant production modes and their respective theoretical dependance are
\begin{itemize}
\item Gluon fusion ($gg \rightarrow h$): $ {\sigma_{GF}}/{\sigma^{\rm SM}_{GF}}= |r_{G}|^2$,
\item Vector boson fusion ($qq\rightarrow h qq$): ${\sigma_{VBF}}/{\sigma^{\rm SM}_{VBF}} = |r_{V}|^2$,
\item Vector boson associated production ($q\bar{q}\rightarrow h V$): ${\sigma_{VH}}/{\sigma^{\rm SM}_{VH}} = |r_{V}|^2$,
\item Top associated production ($gg\rightarrow h t \bar{t}$): $ {\sigma_{h t \bar{t}}}/{\sigma^{\rm SM}_{h t \bar{t}}} =|r_{t}|^2$.
\end{itemize}
In some cases, a channel can include events from several production modes; for instance, the dijet tagged signature in Table~\ref{tab:ratesvalues} is dominantly produced via vector boson fusion, but contains a non-negligible contribution from gluon fusion. The collaborations have made public all the numbers needed to assess the composition in terms of the physical production modes in the references listed in Table~\ref{tab:ratesvalues}. In the fit we use the numbers provided channel by channel, but we find a good uniformity between different final states and the two experiments, giving roughly $75-80\%$ of VBF and $20-25\%$ of GF in the dijet tagged categories. The relative fraction of the untagged mode are close to be the relative cross-section fractions of the different channels. Therefore, where efficiencies are not publicly available ({\it e.g.} ATLAS untagged WW), we use the ratio between cross-sections to determine the composition of the sample. The $VH$ and $ttH$ categories can be taken as pure.

The fits to the data are performed minimizing the $\chi^2$
\be
\chi^2 = \sum_{{\rm i=channels }}{ \frac{(\mu_i(r_j)- \hat{\mu}_i)^2}{\sigma_i^2}}
\ee
where $\mu_i(r_j)$ are predicted rates and $\sigma_i$'s are two-sided errors (depending on the sign of $\mu_i- \hat{\mu}_i$). By performing the $\chi^2$ test we are assuming that the likelihood functions for $\hat{\mu}$ follow an  approximate two-sided gaussian distribution and the correlations can be neglected, both of which have been shown to be valid approximations \cite{Carmi:2012in,Espinosa:2012im}. Furthermore we have compared our results with the two dimensional fits released by the collaborations, obtaining a good agreement.

In the following we refer to  ``preferred'' and ``allowed'' regions. Preferred regions are obtained by varying an $N$-parameters subset of the $r_i$'s, while fixing the other parameters to their SM-values. Strictly speaking, allowed regions are found including all the $r_i$'s, the possibility of flavor non-universality (for instance $r_b \ne r_\tau$), vertex structures different from the SM and the presence of an invisible width. At this stage, with limited precision in the measurements, we find a theoretically inspired (see Sections \ref{sec:predict} and \ref{sec:res}) five dimensional fit a reasonable approximation of the most generic setting. In this framework we obtain lower dimensional confidence intervals
by treating the remaining independent parameters as nuisances (given a lower dimensional point the other parameters are varied to give the best possible fit).

\begin{table}[t!]
\caption{Rates relative to their SM value with the respective 68\% confidence interval, as measured by the ATLAS (top), CMS (middle) and CDF+D0 (bottom) experimental collaborations.}
\begin{center}
\begin{tabular}{|c|c|c|c|c|}
\hline
\textbf{ATLAS} &untagged & dijet tagged & VH & ttH\\
\hline
$h\to\gamma\gamma$ &$1.80\pm0.50$  \cite{ATLASHgg}&$2.7\pm1.3$ \cite{ATLASHgg}&$\times$&$\times$\\
\hline
$h\to WW$  &$1.34^{+0.50}_{-0.59}$ \cite{ATLASHWW}&$\times$&$\times$&$\times$\\
\hline
$h\to ZZ$  & $1.20^{+0.60}_{-0.55}$ \cite{ATLASHZZ}&$\times$&$\times$&$\times$\\
\hline
$h\to \tau\tau$ &(*)$\times$&$0.69_{-0.69}^{+0.71}$\cite{ATLASHtautauSM} &$\times$&$\times$\\
\hline
$h\to b\bar b$ &$\times$&$\times$&$0.7\pm 1.1$ \cite{ATLASHbb}&$\times$\\
\hline
\hline
\textbf{CMS} & untagged & dijet tagged & VH & ttH\\
\hline
$h\to\gamma\gamma$ &  $1.48_{-0.39}^{+0.54}$\cite{CMSHgg} &$2.17_{-0.93}^{+1.4}$  \cite{CMSHgg} &$\times$&$\times$\\
\hline
$h\to WW$  &$0.79_{-0.27}^{+0.26}$ \cite{CMSHWW}&$0.00_{-0.60}^{+0.70}$ \cite{CMSHWW}&$-0.2_{-2.0}^{+2.1}$ \cite{CMSHWW}&$\times$\\
\hline
$h\to ZZ$  &$0.82_{-0.29}^{+0.34}$ \cite{CMSHZZ}&$\times$&$\times$&$\times$\\
\hline
$h\to \tau\tau$ &$0.8_{-1.2}^{+1.4}$ \cite{CMSHtautauSM}&$0.82_{-0.75}^{+0.83}$ \cite{CMSHtautauSM}&$1.1_{-1.7}^{+2.0}$ \cite{CMSVHtautau}&$\times$\\
\hline
$h\to b\bar b$ &$\times$&$\times$&$1.33_{-0.65}^{+0.68}$ \cite{CMSVHbb}&$-0.8_{-2.0}^{+2.2}$ \cite{CMSttHbb}\\
\hline
\hline
\textbf{CDF+D0} &untagged & dijet tagged & VH & ttH\\
\hline
$h\to b\bar b$ &$\times$&$\times$&$1.61^{+0.74}_{-0.75}$зк&$\times$\\
\hline
\end{tabular}
\end{center}
(*) We have not included the non-VBF $\tau\tau$ categories for ATLAS. The information on the cross correlation is not provided making difficult to asses the contamination of the combined result. To extract the VBF only rate from the $\mathrm{CL_s}$ plot we follow the procedure described in~\cite{Giardino:2012ww}.
\label{tab:ratesvalues}
\end{table}%

\section{Predictions and Results}

In this Section we discuss in more detail the predictions of natural supersymmetry and compare them with the results of the fit to Higgs data. We consider four simplified scenarios: a type-II 2HDM , a  type-II 2HDM with a SM singlet, light stops and light charginos  both in the Higgs decoupling limit. At the end of the section we discuss the more general five parameters fit.

\label{sec:res}
\subsection{The type-II Two Higgs Doublet Model}\label{sec:2hdm}
If the superpartners are relatively heavy, i.e. $m_{\widetilde t_1}\gtrsim 500$~GeV and $m_{\chi_1^\pm}\gtrsim 200$~GeV, loops corrections to Higgs couplings are at most of the order of 10$\%$ in a broad region of parameter space, provided that the stop mixing is limited by naturalness~\cite{Blum:2012ii}. Therefore it is easy to decouple these effects without incurring excessive fine-tuning and without the need to single out narrow corners of parameter space.

With this in mind, we begin by exploring the tree-level corrections in the MSSM and its extensions, using the generic type-II 2HDM scalar potential and Yukawa couplings~\cite{Gunion:2002zf, BRanco:2011iw},
\ba\label{eq:V2hdm}-\mathcal{L}&=&
m_1^2|H_d|^2+m_2^2|H_u|^2\nn\\
&+&\frac{\lambda_1}{2}|H_d|^4+\frac{\lambda_2}{2}|H_u|^4+\lambda_3|H_d|^2|H_u|^2+\lambda_4|H_d^\dag  H_u|^2\nn\\
&+&\frac{\lambda_5}{2}(H_d\cdot H_u)^2+(H_d \cdot H_u)\left(m_{12}^2+\lambda_6|H_d|^2+\lambda_7|H_u|^2\right)+h.c.\nn\\
&+&Y_t\,H_u \cdot Q_3 \,t_R^c+Y_b\,H_d \cdot Q_3 \,b_R^c+Y_\tau \,H_d \cdot L_3 \,\tau_R^c+h.c.\,,\ea
where $A \cdot B \equiv \epsilon_{ij} A^i B^j$, and we adopt the usual definition $H_u=(H_u^+,~H_u^0)^T$ and $H_d=(H_d^0,~H_d^-)^T$. We also take all couplings in the Higgs potential to be CP-conserving. In this setting the Higgs couplings to the SM particles depend on two parameters, the ratio of Higgs vevs~\cite{Gunion:2002zf}
\be\tan\beta=\frac{\langle H_u\rangle}{\langle H_d\rangle},\ee
and the mixing angle of the CP-even Higgs states, defined by
\be
\left(
\begin{array}{c}
 H \\
 h \\
\end{array}
\right)=\sqrt{2}\left(
\begin{array}{cc}
 \cos\alpha  & \sin\alpha  \\
 -\sin\alpha  & \cos\alpha \\
\end{array}
\right)\left(
\begin{array}{c}
 \text{Re}~H_d^0 \\
 \text{Re}~H_u^0 \\
\end{array}
\right).
\ee
where $h$ is the observed $126$ GeV state. At tree-level, $r_b=r_\tau$,
\be\label{eq:rs2hdm}
r_b=-\frac{\sin\alpha}{\cos\beta},\;\;\;r_t=\frac{\cos\alpha}{\sin\beta},\;\;\;r_V=\sin\left(\beta-\alpha\right),
\ee
and in the MSSM
\be \label{eq:maab}
\frac{\sin 2\alpha}{\sin 2\beta} = - \frac{m_H^2 +m_h^2}{m_H^2 - m_h^2},\;\;\;\;\;\;\;\;\;\; \frac{\tan 2\alpha}{\tan 2\beta} =  \frac{m_A^2 +m_Z^2}{m_A^2 - m_Z^2}\, .
\ee
Without loss of generality, we can take $\tan\beta >0$~\cite{Carena:2002es}, with Eq. (\ref{eq:maab}) implying $\sin\alpha <0$. We require that the top Yukawa does not blow up above the electroweak scale, which imposes the lower bound $\tan\beta\geq1$ \cite{Martin:1997ns}. As discussed in Section~\ref{sec:predict}, demanding loop corrections to the type-II 2HDM structure to be negligible provides the additional constraint $\tan\beta \leq 40$ (that can be relaxed or made stronger depending on the level of fine-tuning that we allow in the theory).

\begin{figure}[t!]
\begin{center}
\includegraphics[width=\textwidth]{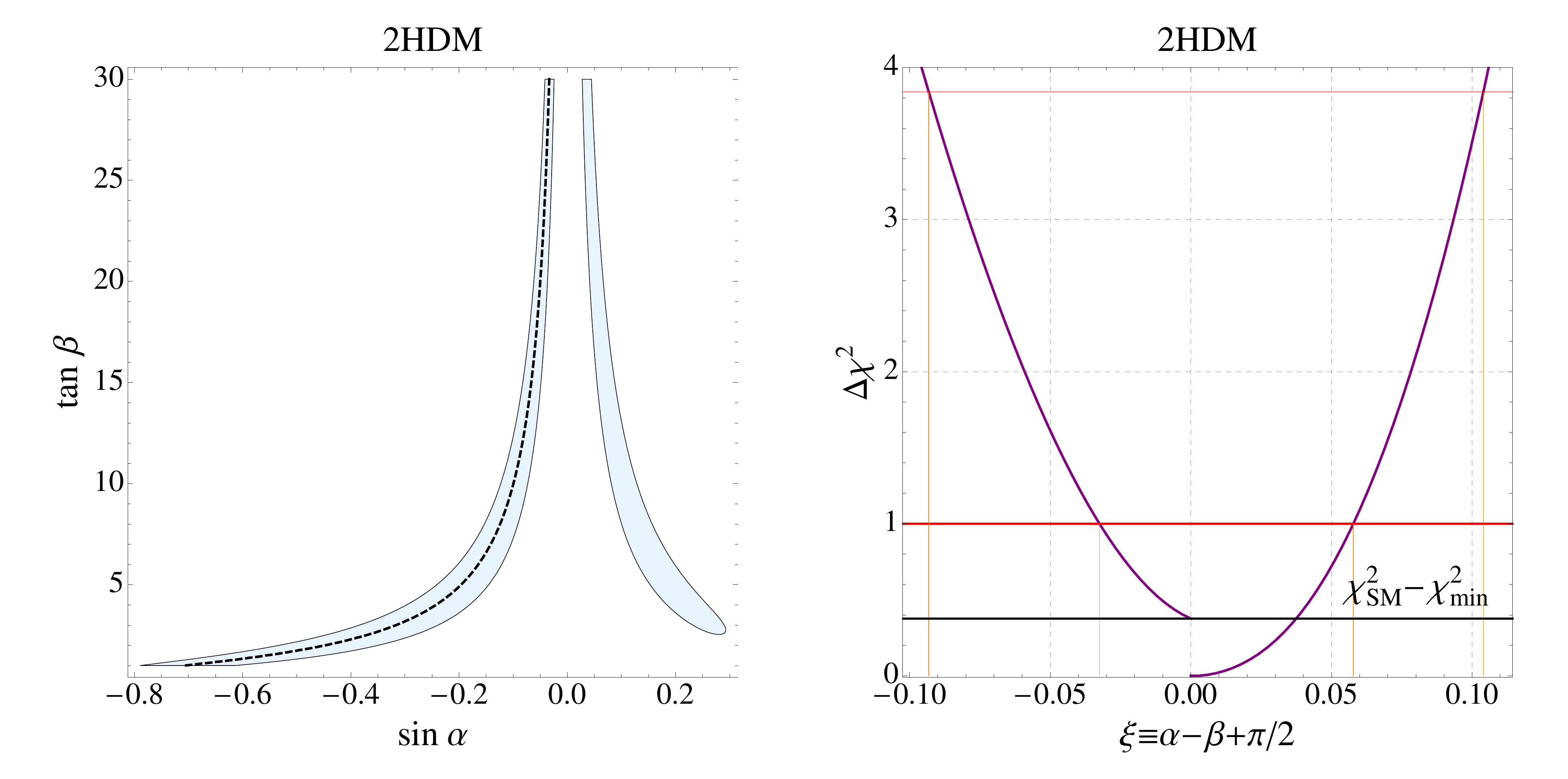}
\end{center}
\caption{{\bf Left:} $\chi^2$ contours corresponding to the $95$\% confidence level in the $(\sin\alpha, \tan\beta)$ plane (for a type-II 2HDM). Note that in the MSSM $\sin\alpha <0$. The black dashed line corresponds to the decoupling limit of the MSSM $\alpha=\beta-\pi/2$.
{\bf Right:} $\Delta \chi^2$ vs. $\xi(\equiv \alpha -\beta +\pi/2)$. The red lines mark the $68\%$ and $95\%$ confidence levels heights of the $\Delta \chi^2$. In black the offset between the SM value of the $\chi^2$ and the 2HDM value at the minimum. The discontinuity in the $\chi^2$ is physical and it is illustrated by Eq.~(\ref{eq:approxXi}). Note that in the tree-level MSSM $\xi \geq 0$.}
\label{fig:rbrt}
\end{figure}

Since the measured rates broadly agree with the SM, the results of the fit, which are shown in Figure \ref{fig:rbrt}, point to the decoupling limit of the model, $\xi  \equiv \alpha -\beta +\pi/2 \approx 0$. This translates into the bound $|\xi| \lesssim 0.1$ at $95\%$ C.L. Note that the discontinuity in the $\chi^2$ is physical and can be easily understood by expanding Higgs couplings for small $\xi$
\be
\label{eq:approxXi}
r_b=1-\xi \tan\beta+\mathcal{O}(\xi^2), \quad r_t=1+\xi/\tan\beta+\mathcal{O}(\xi^2), \quad r_V=1-\frac{\xi^2}{2}+\mathcal{O}(\xi^4)\, .
\ee
Even for extremely small values of $\xi$ the correction to $r_b$ can be significant for large $\tan\beta$ and depends on the sign of $\xi$. The one dimensional $\chi^2$ was obtained by treating $\tan\beta$ as a nuisance, so the profiling is selecting a $\tan\beta$ that can contribute an observable deviation on the side where $r_b$ is depleted, while for $\xi<0$ a $\tan\beta$ giving the smallest possible deviation is singled out. For this reason the offset between the two sides is the same as the offset between the deepest minimum and the SM.

The result, albeit disappointing from the perspective of finding new physics, allows us to consider a particular realization of the decoupling limit in which more interesting statements can be made on the mass of the second Higgs doublet. In general it is difficult to translate the fit results for $\alpha$ and $\beta$ into a direct mass exclusion or into an exclusion on the Lagrangian parameters. However, if the 2HDM potential respects an approximate PQ symmetry, or a $Z_2$: $H_d \rightarrow - H_d , \; H_u\to H_u$, advances can be made.

Assuming that the second Higgs doublet is parametrically heavier than $H_u$, we can expand the light Higgs couplings in the $Z_2$ breaking spurion $B/M_1^2$, where\footnote{In the following we loosely refer to this expansion as arising from an approximate PQ symmetry, but $\lambda_5$ can be large without spoiling our approximation and preserving only the  $Z_2$: $H_d \rightarrow - H_d , \; H_u\to H_u$.}
\be
M_1^2=m_1^2+\frac{\lambda_{35}\left(H_u^0\right)^2}{2}\,,\;\;\;B=m_{12}^2+\frac{\lambda_7\left(H_u^0\right)^2}{2}\, .
\ee
and $\lambda_{35} = \lambda_3 + \lambda_5$.
Up to leading order $\langle B/M_1 \rangle = 1/\tan\beta $, thus the expansion converges rather fast in $1/\tan\beta$, {\it e.g.}, only a $\mathcal{O}(10\%)$ error for $\tan\beta= 3$. This setting is extremely predictive, giving (in the MSSM and many of its extensions where $\lambda_7=0$)~\cite{Blum:2012kn}
\be\begin{array}{rcl}
r_b&=&\left(1-\dfrac{m_h^2}{m_H^2}\right)^{-1}\left(1-\dfrac{\lambda_{35}v^2}{m_H^2-m_h^2}\right)
\times\left\{1+\mathcal{O}\left(\dfrac{1}{\tan^2\beta}\right)\right\} \, , \\
r_t &=& r_V = 1+ \mathcal{O}\left(1/\tan^2\beta\right)\, .
\end{array}\label{eq:rbexp}\ee
In this limit, the corrections to $r_b$ depend on a single combination of quartics, $\lambda_{35}$, and the mass of the heavy scalar Higgs $m_H$. Note that if we assume the approximate PQ and some degree of custodial $SU(2)$ symmetry, the differences in the masses of the heavy Higgses are small~\cite{Dine:2007xi}: $m_H \simeq m_A \simeq m_{H^+}$. Thus, to first approximation $m_H$ can be considered the mass of the full heavy doublet.

\begin{figure}[t!]
\begin{center}
\includegraphics[width=0.45\textwidth]{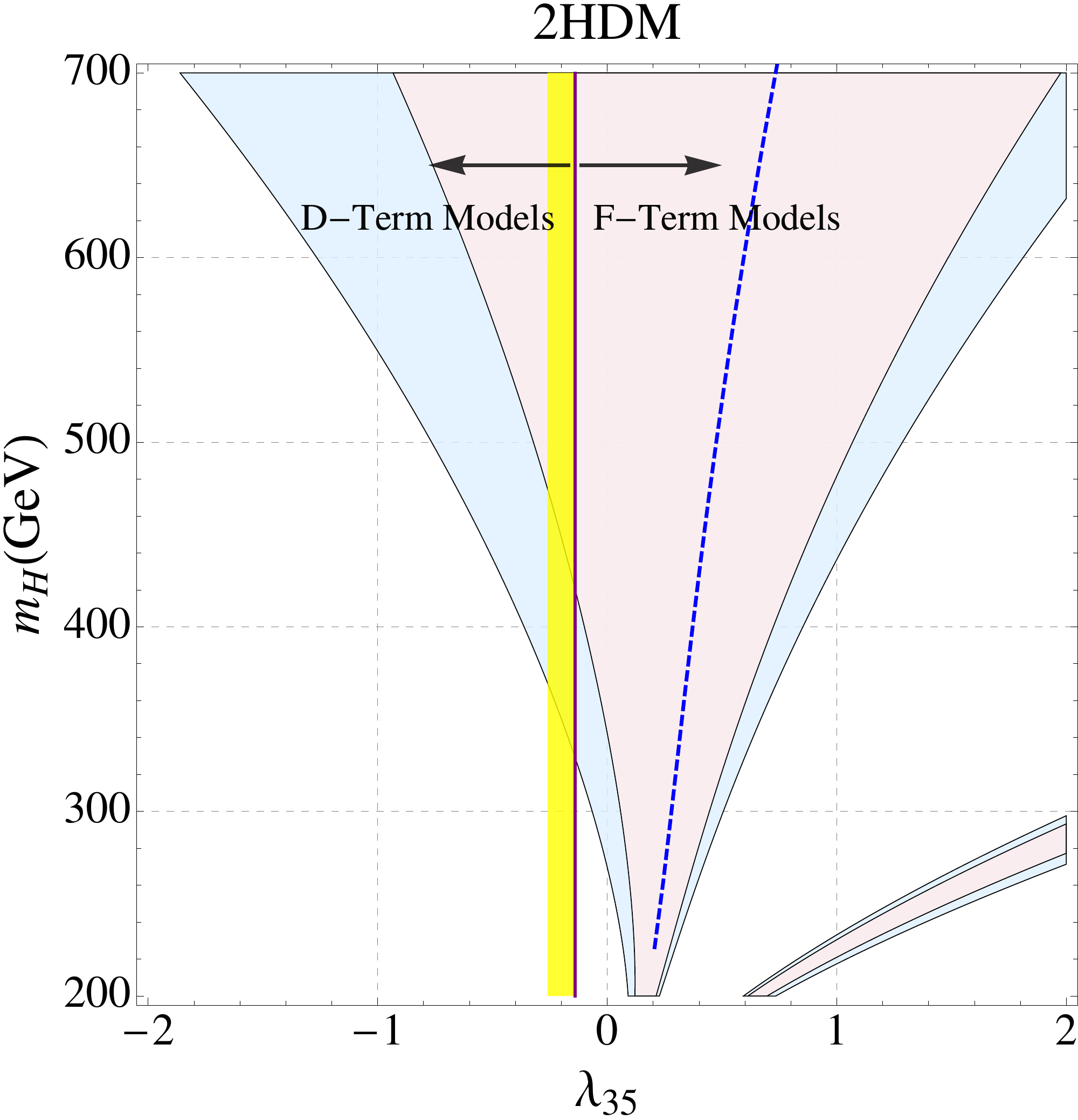}\quad
\end{center}
\caption{{\bf Left:} $\chi^2$ contours corresponding to the $68$\% and $95$\% confidence levels in the $(\lambda_{35}, m_H)$ plane, where $r_b$, $r_t$ and $r_V$ are given by \ref{eq:rbexp}. The blue dashed line corresponds to the best fit points, the solid purple line to the tree-level value of $\lambda_{35}$ in the MSSM. The band in yellow covers to the possible values of $\lambda_{35}$ in pure D-term models (defined in section~\ref{sec:res}). }
\label{fig:mssm2hdm}
\end{figure}

The result of the fit is depicted in Figure~\ref{fig:mssm2hdm}. In the MSSM, $\lambda_3 = -(g^2+(g')^2)/4\approx -0.14$ and $\lambda_5=0$ at tree-level, while the loop corrections are small. This leads to a preferred region at 95\% C.L. with $m_H\gtrsim370$~GeV. However the $\tan\beta$ expansion clearly introduces an error that in the MSSM goes in the direction of relaxing our bound\footnote{
The approximation in (\ref{eq:rbtb2}) is valid only at tree-level and the leading term can be obtained from Eq. (\ref{eq:rbexp}) ignoring the loop corrections to $\lambda_2$: $m_h \to m_Z\left(1+1/\tan^2\beta\right)$.}
\be\label{eq:rbtb2}
r_b^2 \approx 1+\frac{4 m_Z^2}{m_A^2}-\frac{12 m_Z^2}{m_A^2}\frac{1}{\tan^2\beta}+\mathcal{O}\left(\frac{m_Z^4}{m_A^4}, \frac{1}{\tan^4\beta}\right)\, .
\ee
In the following section we obtain a limit on $m_A$ as a function of $\tan\beta$ using the full tree-level expression in the MSSM together with the most important loop corrections. The effect of taking into account the $\mathcal{O}(1/\tan^2\beta)$ variations of the other couplings is much smaller and will be commented upon while repeating this exercise in the five dimensional fit case and in the MSSM with the full $\tan\beta$ dependence.

This expansion is also relevant to a vast class of models in which the supersymmetric Higgs sector is not necessarily minimal, but it is still well described by a type-II 2HDM at low energy. This is the case for $D$-term models raising the Higgs mass via extra gauge interactions and also for some areas of the parameter space of $F$-term models.

We define $D$-term models as the class of theories in which the two Higgs doublets are charged under new gauge interactions. We will restrict to gauge theories that do not forbid the $\mu$ term, otherwise additional $F$-terms should be present to generate it. In $D$-term models the $b\bar b$ rate is generically enhanced. This descends from the fact that if the charge assignment allows the presence of a $\mu$ term, the new $D$-term contributes to the relevant part of the scalar potential
\ba \label{eq:D-potential}
\left(\left|H_u^0\right|^2-\left|H_d^0\right|^2+ \xi_D \right)^2,
\ea
with some positive definite coefficients in front. $\xi_D$ contains both gauge symmetry breaking terms that can be absorbed by the soft masses $m_1^2$ and $m_2^2$ and new fields charged under the gauge groups. Thus the new effect will be a contribution to $\lambda_3$ that is always negative, resulting in stronger bounds on $m_H$.  This is illustrated in Figure~\ref{fig:mssm2hdm}, by the yellow band corresponding to variation of the $D$-term contribution to the Higgs mass from $0$ to $m_h-m_Z$. Note that to obtain an observable effect from the extra gauge sector, it is necessary to have a large supersymmetry breaking mass for the new gauge bosons. If the heavy gauge bosons are integrated out supersymmetrically, there is a shift in the quartic $\delta \lambda_3 \sim \mu^2/M_V^2$~\cite{Batra:2003nj}, where $\mu$ is limited by naturalness while $M_V \gtrsim 3$~TeV due to EWPTs~\cite{Chivukula:2003wj}.

For $F$-term models the applicability of the approximation in Eq. (\ref{eq:rbexp}) is limited, but we can consider a superpotential of the form
\be
\mathcal{W}=\lambda S H_u H_d + f(S)\, ,
\label{eq:NMSSMW}
\ee
where $S$ is either a singlet under the SM gauge groups or an $SU(2)$ triplet with $Y=0$. Integrating out $S$ and neglecting $v^2/m_s^2$ effects (where $m_s$ is a supersymmetry breaking mass), the leading observable modification to the Higgs potential is
\be
\lambda_3^{\mathrm{MSSM}}\to \lambda_3^{\mathrm{MSSM}} +|\lambda|^2\, ,
\ee
which again shifts $\lambda_3$ by a definite sign, but in the opposite direction with respect to $D$-term models. Note that in this case small $\tan\beta$ may be preferred to lift the tree-level Higgs mass,
\be
m_h^2 \leq m_Z^2 \left(\cos^2 2\beta + \frac{2|\lambda|^2}{g^2+\left(g'\right)^2} \sin^2 2\beta\right)\,
\ee
and the expansion given in Eq. (\ref{eq:rbexp}) may not be theoretically motivated. Furthermore to leading order, in gauge and gaugino mediation, the singlet supersymmetry breaking mass $m_s$ in the NMSSM vanishes at the mediation scale~\cite{Morrissey:2008gm}, so it is necessary to introduce some deformations of the simplest scenarios to lift the mass above the other soft terms at the electroweak scale. There are nonetheless large areas of parameter space, in which the expansion is reliable, that give the correct Higgs mass. For instance, including stop corrections with $m_{\tilde t_1} \approx m_{\tilde t_2} = 380$ GeV and $X_t=0$, we get $m_h=126$ GeV for all values of $\lambda$ between 1 and 2, varying $\tan\beta$ between 3 and 8~\cite{Blum:2012ii}. Considering this range for $\lambda$ would correspond to taking $\lambda_{35}$ between 1 and 4, thus giving a favored region, that for $\lambda=1$, is already $m_H \gtrsim 460$ GeV at 95\% C.L.

\subsection{Heavy Higgses and $\mathbf{ m_h=126}$ GeV in the MSSM}\label{sec:hmssm}
As discussed above only two parameters are needed in the MSSM, at tree-level, to specify all Higgs couplings. Therefore we can express the fit constraints in the ($m_A$, $\tan\beta$) plane and compare our results with direct searches. To do so, the requirement of naturalness needs to be abandoned to allow for the stops' radiative corrections to raise the Higgs mass to $126$ GeV. Considering only the leading loop correction (which is the only one that survives in the limit $\mu\to 0$) only the $H_u-H_u$ element of the CP even mass matrix receives a correction. The tree-level relations in the previous section are correspondingly shifted to~\cite{Djouadi:2005gj}
\be
\begin{array}{rcl}
\dfrac{\tan 2 \alpha}{\tan 2 \beta}&=&\frac{m_A^2+m_Z^2}{m_A^2-m_Z^2+\delta m^2/\cos 2\beta}\, ,\\
m_H^2 &=& -m_h^2+m_A^2+m_Z^2+\delta m^2\, ,
\end{array}
\ee
where $\delta m^2$ can be approximated by the leading stop correction
\be
\delta m^2 \approx \frac{3 m_t^4}{2\pi^2 v^2 \sin^2 \beta}\left[\log \frac{M_{\tilde t}^2}{m_t^2}+\frac{X_t^2}{2 M_{\tilde t}^2}\left(1-\frac{X_t^2}{6 M_{\tilde t}^2}\right)\right], \quad M_{\tilde t}=\frac{1}{2}\left(m_{\tilde t_1}+m_{\tilde t_2}\right)\, ,
\ee
but in the following we always fix $\delta m^2$ to the $m_A$ and $\tan\beta$ dependent value giving the correct Higgs mass
\be
\delta m^2|_{m_A, \tan\beta} = \dfrac{m_h^2 (m_A^2 - m_h^2 + m_Z^2) - m_A^2 m_Z^2 \cos^2 2 \beta}{
 m_Z^2 \cos^2\beta + m_A^2 \sin^2\beta-m_h^2}\, .
\ee

\begin{figure}[t!]
\begin{center}
\includegraphics[width=0.7\textwidth]{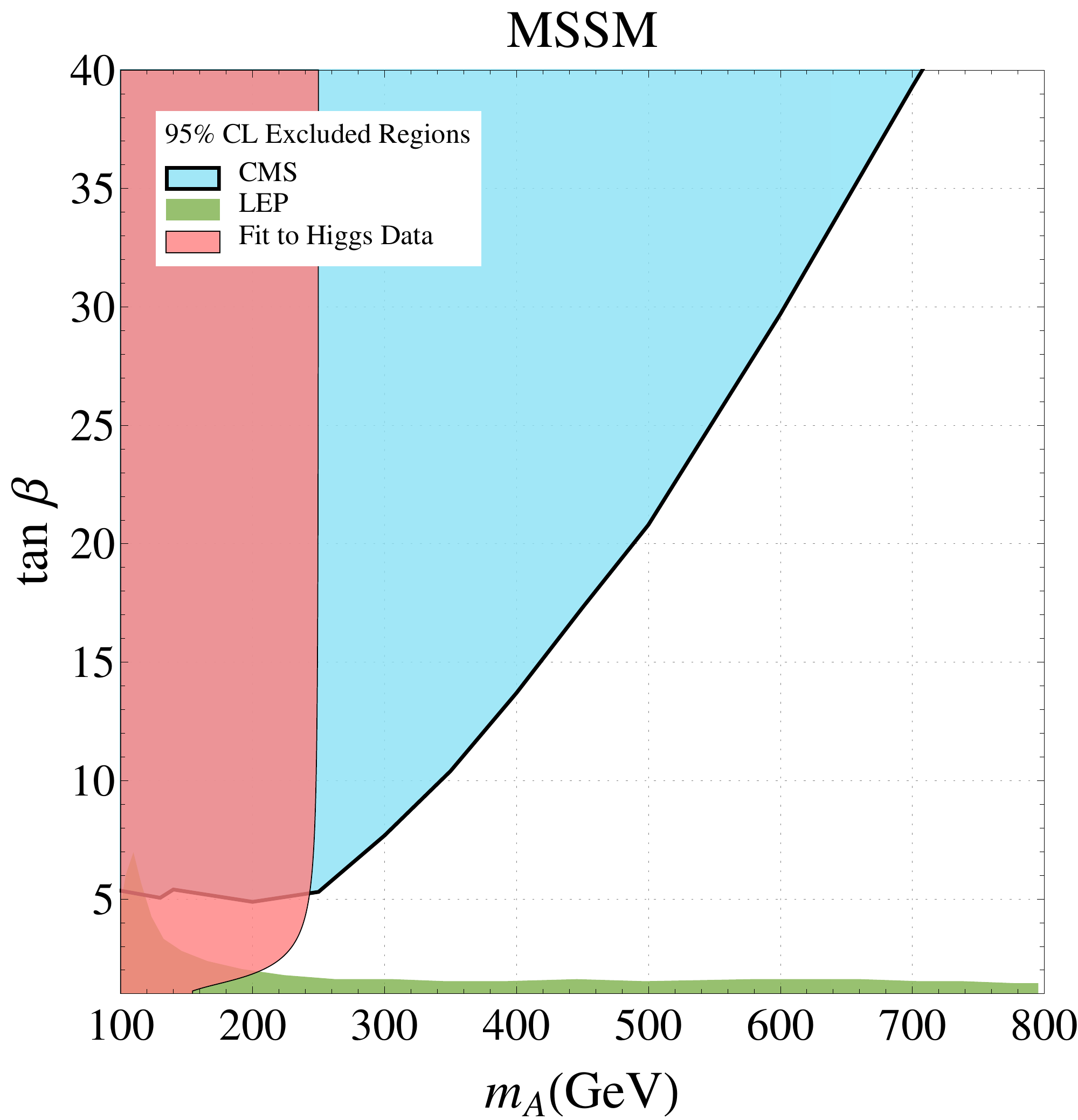}
\end{center}
\caption{Comparison between the fit exclusion in the $(m_A, \tan\beta)$ plane and the direct exclusions from CMS \cite{CMSMSSMHtautau} and LEP~\cite{Schael:2006cr}. The excluded region was obtained profiling the full five dimensional $\chi^2$. Note that the experimental collaborations use the $m_h^{\mathrm{max}}$ scenario~\cite{Carena:2002qg} to set their limits, without imposing $m_h \approx 126$ GeV. Varying the soft supersymmetry breaking parameters might lead to modifications of the observed bounds~\cite{Carena:2005ek}.}
\label{fig:maTb}
\end{figure}%

Before describing the results of the fit we review current LHC searches for MSSM Higgses. Direct searches for the neutral MSSM Higgses are currently performed in the $\phi\to\tau^+\tau^-$~\cite{CMSMSSMHtautau, ATLASHtautauMSSM}, $b\bar b \phi\to b\bar b\mu\mu$~\cite{CMSHbbmumu} and $b\bar b \phi\to b\bar b b \bar b$~\cite{CMSHbbbb} channels. The strongest bound is set by the recently updated CMS measurement~\cite{CMSMSSMHtautau} and ranges from $m_H \gtrsim 250$ GeV for $\tan\beta=5$ to $m_H \gtrsim 700$ GeV for $\tan\beta=40$. The sensitivity vanishes below $\tan\beta=5$. These searches benefit from large $\tan\beta$, both thanks to new production mechanisms that become important ($b$ and $b\bar b$ associated production for instance) and from the increase in the branching ratio to $\tau^+\tau^-$. We also include the LEP bound~\cite{Schael:2006cr} in our comparison with Higgs rates, but we do not consider the implications of searches for the charged Higgs that are currently not as sensitive as the $\phi \to \tau^+\tau^-$ one plus the LEP constraint.
The experimental collaborations use the $m_h^{\mathrm{max}}$ scenario~\cite{Carena:2002qg} to set their limits, and varying the soft supersymmetry breaking parameters might lead to modifications of the observed bounds~\cite{Carena:2005ek}. The effects are stronger for large $\tan\beta$ mainly due to loop corrections to $y_b$ and can not produce any significant gain in sensitivity in the region $3\lesssim\tan\beta\lesssim 12$, where production cross sections become too small and our analysis starts to be competitive. Additionally, the collaborations do not impose $m_h \approx 126$ GeV. Ideally, the bounds from ~\cite{CMSMSSMHtautau} should be reinterpreted as a bound on $\sigma (pp\rightarrow \phi)Br(\phi \rightarrow \tau^+ \tau^-)$, but it is beyond the scope of this work.

Figure~\ref{fig:maTb} shows the overlay of these bounds with the results of the fit to the Higgs data. In the slice $2\lesssim\tan\beta\lesssim 5$ Higgs rates are probing regions of the parameter space not directly accessible to CMS and LEP. The fit results in Figure~\ref{fig:maTb} are an anticipation of the five dimensional fit discussed in section~\ref{sec:4D} since the excluded region was obtained profiling the five dimensional $\chi^2$, treating $r_\gamma$ and $r_G$ as nuisances. This is reflected in the fact that for large $\tan\beta$ there is a lower bound $m_A \gtrsim 250$~GeV, in agreement with the result of the PQ expansion in Figure~\ref{fig:mssm2hdm4D}.

\subsection{Singlet Mixing}\label{sec:mixing}
\begin{figure}[t!]
\begin{center}
\includegraphics[width=0.55\textwidth]{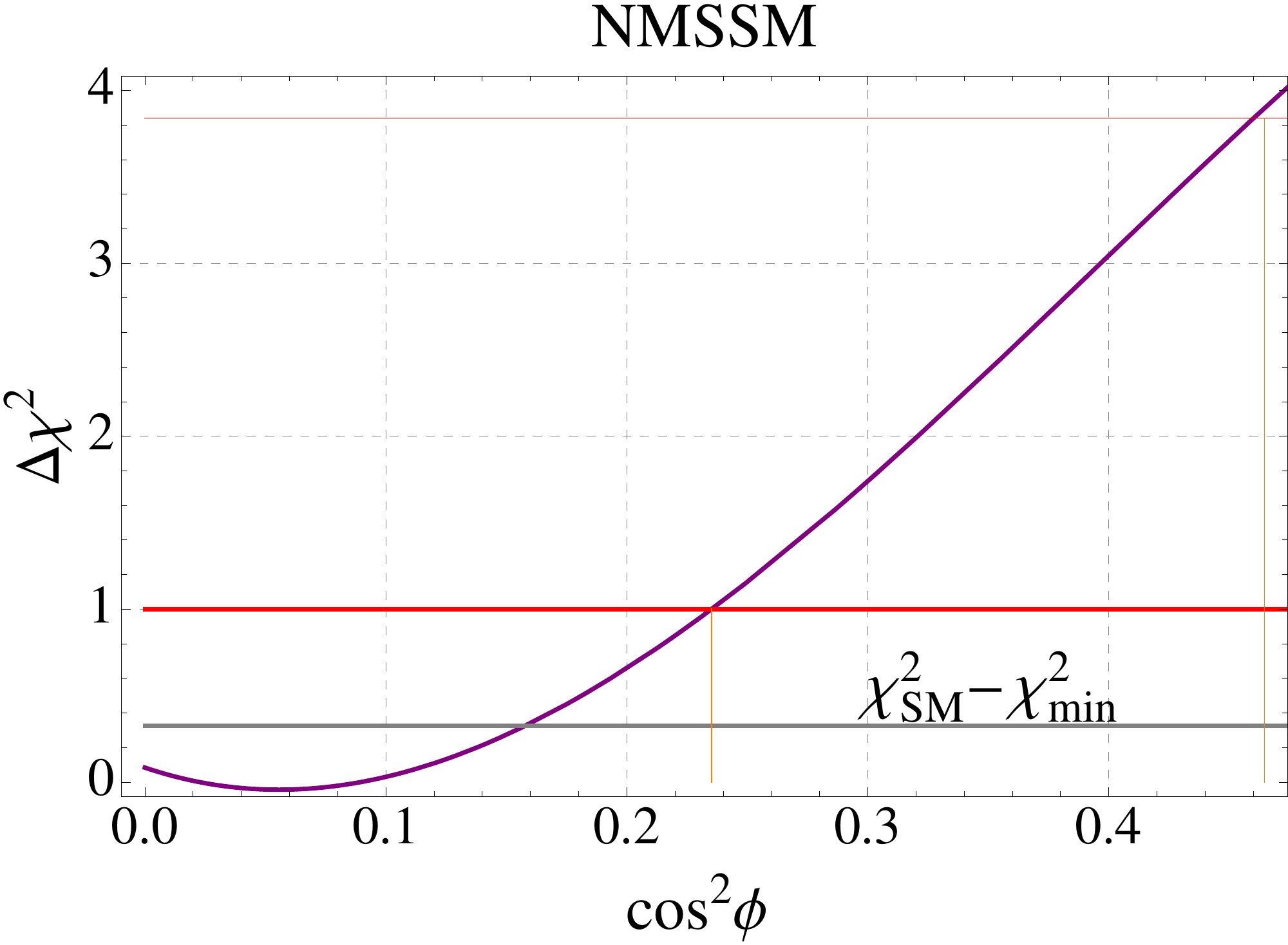}
\end{center}
\caption{$\Delta\chi^2$ as a function of the singlet mixing parameter $\cos^2\phi$. The $68$\% and $95$\% confidence levels are marked in red. In black we show the offset between the $\chi^2$ at the minimum and the SM minimum.}
\label{fig:nmssm}
\end{figure}

Finally, we consider the effects of the singlet mixing with the two Higgs doublets. We take the superpotential (\ref{eq:NMSSMW}) and do not specify the detailed form of the soft supersymmetry breaking terms or extra superpotential interactions. For simplicity we take all new couplings to be CP conserving and assume the SM-like Higgs to be CP even and the lightest Higgs state in the theory. With these assumptions we do not need to consider exotic decays that could be triggered by approximate $U(1)$'s, including the $U(1)_R$, generating light pseudo-goldstone bosons in the Higgs sector. This considerably simplifies the problem, leaving the relevant parameters to be contained in the $3\times 3$ mixing matrix for the scalar CP even states,
\be
\label{eq:matrix}
\left(
\begin{array}{c}
 H_2 \\
 H_1 \\
 h \\
\end{array}
\right) = \left(
\begin{array}{ccc}
 s_{\phi } c_{\chi } & -c_{\alpha } s_{\chi }+s_{\alpha } c_{\phi } c_{\chi } & -s_{\alpha } s_{\chi }-c_{\alpha } c_{\phi } c_{\chi } \\
 s_{\phi } s_{\chi } & c_{\alpha } c_{\chi }+s_{\alpha } c_{\phi } s_{\chi } & s_{\alpha } c_{\chi }-c_{\alpha } c_{\phi } s_{\chi } \\
 c_{\phi } & -s_{\phi } s_{\alpha } & s_{\phi } c_{\alpha } \\
\end{array}
\right) \left(
\begin{array}{c}
 S \\
 H_d \\
 H_u \\
\end{array}
\right)\, ,
\ee
where $c_x = \cos x, s_x = \sin x$. In this notation, the tree-level Higgs couplings depend on $(\alpha, \phi, \beta)$, and  Eq. (\ref{eq:rs2hdm}) is modified to
\be\label{eq:rs2hdm}
r_b=-\sin\phi \frac{\sin\alpha}{\cos\beta},\;\;\;r_t=\sin\phi \frac{\cos\alpha}{\sin\beta},\;\;\;r_V=\sin\phi\sin\left(\beta-\alpha\right),
\ee
where $\cos\phi \equiv \langle S | h \rangle$ measures the amount of singlet in the lightest Higgs. The relations (\ref{eq:maab}) are only valid in the MSSM, so that $m_A$ is no longer determined solely (at tree level) from the mixing parameters $\alpha$ and $\beta$. Thus, we do not impose any constraint on $\alpha$ and $\phi$, but still restrict to $\tan\beta \ge 1$. In Fig.~\ref{fig:nmssm}, we plot $\Delta \chi^2 $ vs $\cos^2\phi$. Since the presence of singlet mixing will only decrease the observed rates, there is no improvement in the fits over the SM and we obtain an upper bound $|\cos\phi| \lesssim 0.7$ at $95\%$ C.L. It is worth pointing out that even though this bound was derived assuming only tree-level effects are present, it is similar to the bound obtained in the full 5D space, as will be discussed in Section \ref{sec:4D}.

\subsection{Stops}\label{sec:stops}
LHC searches, together with naturalness, have stimulated a vast offspring of models of mediation that produce light stops, decoupling all other squarks~\cite{Dimopoulos:1995mi, Cohen:1996vb, Auzzi:2011eu, Csaki:2012fh, Craig:2012di, Craig:2012hc}. In addition, bottom-up approaches, driven mainly by FCNC constraints, still allow stops to be fairly light~\cite{Cohen:1996vb, Barbieri:2009ev, Barbieri:2010pd}. It is then worth exploring the stop plane in view of Higgs data.

Using the Higgs low energy theorem~\cite{Kniehl:1995tn, Ellis:1975ap, Shifman:1979eb} it is straightforward to obtain the change in the gluon fusion rate from integrating out the stops (neglecting $D$-terms)
\be
\delta r_G^{\tilde t}\approx \frac{m_t^2}{4}\left[\frac{1}{m_{\tilde t_1}^2}+\frac{1}{m_{\tilde t_2}^2}-\frac{X_t^2}{m_{\tilde t_1}^2m_{\tilde t_2}^2}\right]\, . \label{eq:rGstops}
\ee
where $X_t = A_t - \mu\cot\beta$ and $m_{\tilde t_{1}}^2<m_{\tilde t_{2}}^2$ are the two eigenvalues of the stop mass matrix
\be
\mathcal{M}^2_{\tilde t}=\left(\begin{array}{cc}m_{\tilde Q}^2+m_t^2+\left(\frac{1}{2}-\frac{2}{3}s_w^2\right)m_Z^2 \cos 2\beta & m_t X_t \\ m_t X_t & m_{\tilde u}^2+m_t^2+\frac{2}{3}s_w^2m_Z^2 \cos 2\beta \end{array}\right)\, .
\ee
The approximation in (\ref{eq:rGstops}) is valid up to order $m_h^2/4 m_{\tilde t_1}^2$, but is useful to give a qualitative picture of the effect. Nevertheless all the plots and the numbers that are quoted here are obtained with the full MSSM one loop result~\cite{Djouadi:2005gj}.

\begin{figure}[t!]
\begin{center}
\includegraphics[width=1\textwidth]{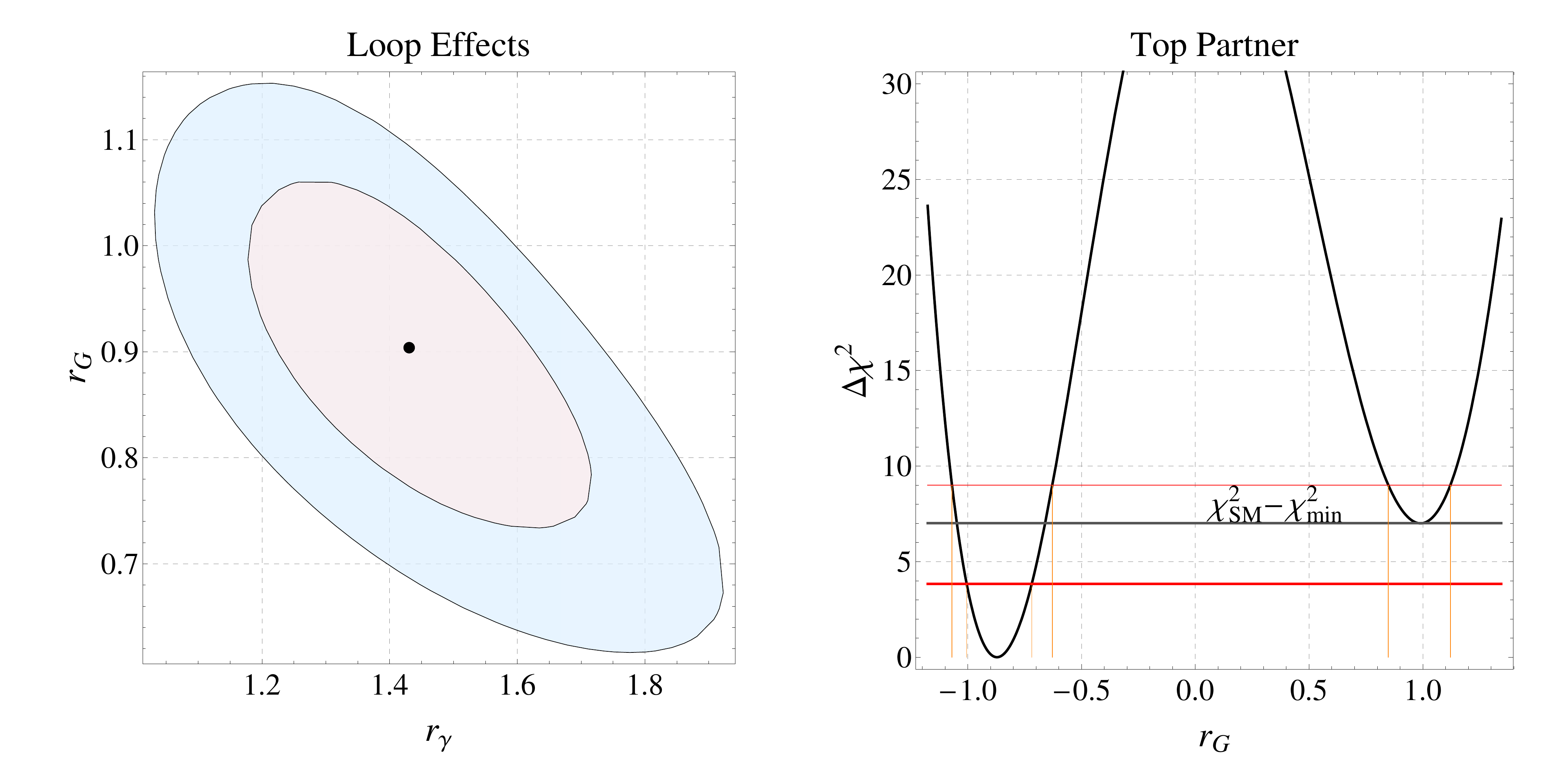}
\end{center}
\caption{{\bf Right:} $\chi^2$ contours at the $68$\% and $95$\% confidence levels in the $(r_G, r_\gamma)$ plane.  All other parameters are fixed to their SM value. {\bf Left:} $\Delta \chi^2$ vs. $r_G$ for the stops fit, with $95\%$ and $99.73\%$ heights marked in red. In gray we show the offset between the $\chi^2$ at the minimum and the SM minimum.}
\label{fig:loops}
\end{figure}%

\begin{figure}[t!]
\begin{center}
\includegraphics[width=\textwidth]{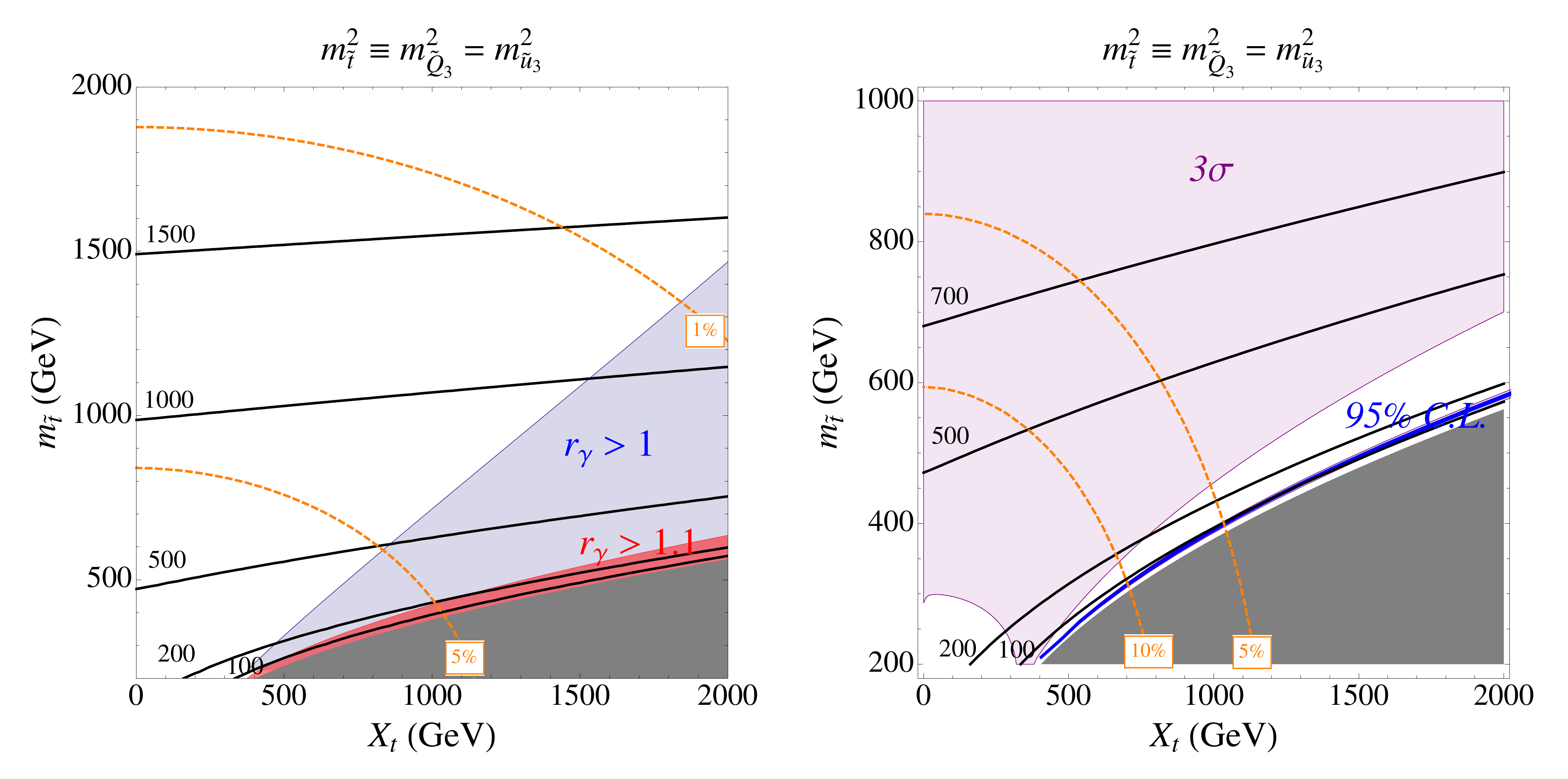} \quad
\end{center}
\caption{{\bf Left:} Regions in the stop plane ($m_{\tilde t}\equiv m_{\tilde Q_3}=m_{\tilde u_3}$) where the $h\to\gamma\gamma$ rate is enhanced with respect to the SM. Black contours give the lightest stop mass in GeV. In orange percent fine-tuning ($\Delta^{-1}$) contours, as defined in equation~\ref{eq:ft}. {\bf Right:} Shaded areas correspond to $95\%$ (blue) and $99.73\%$ (light purple) confidence levels in the $(X_t, m_{\tilde t})$ plane. The black lines are contours of constant $m_{\tilde t_1}$ in GeV. In orange percent fine-tuning ($\Delta^{-1}$) contours are shown, as defined in Eq.~\ref{eq:ft}. In both panels the gray area corresponds to the lightest stop becoming tachyonic.}
\label{fig:stopsTheo}
\end{figure}

The $h\gamma\gamma$ vertex correction can be computed rescaling the contribution to the gluon-gluon amplitude with the appropriate factors of $N_c$, $Q^2$ and $t_c$ (the Dynkin index of the color representation). Since the SM $h\to\gamma\gamma$ loop is dominated by the $W$ boson contribution, the two corrections are opposite in sign, with roughly a factor of $4$ difference in magnitude
\be
r_G = 1+ \delta r_{G}^{\tilde t} ~~~~~~~~~ r_\gamma \approx 1 - 0.27 \delta r_{G}^{\tilde t}. \label{eq:delta}
\ee
The current data point to $|r_G| \approx 0.9$ and $|r_\gamma| \approx 1.4$ as can be seen in the left panel of Figure~\ref{fig:loops}. This can be realized only in a small corridor of the stop plane depicted in the left panel of Figure~\ref{fig:stopsTheo}, that is mostly outside of the $10\%$ tuning region and requires the presence of a very light stop $m_{\tilde t_1} \lesssim 200$~GeV\footnote{Allowing for larger values of $X_t$ it is possible to accommodate large $h\to \gamma\gamma$ enhancements also with an heavier stop, for instance we could have $A_t\approx 10$ TeV, $m_{\tilde t_1}\approx 400$~GeV and still $\delta r_\gamma^{\tilde t}\approx 0.5$~\cite{Buckley:2012em}.}. In this corner of parameter space, the stops' contribution to the gluon fusion rate is opposite in sign and roughly double in magnitude with respect to the top quark. This region is  not only tuned by more than $1\%$ if we demand that the gluino be heavier than 1 TeV, but can also generate color breaking minima~\cite{Reece:2012gi}.

The preferred area with $|r_\gamma| > 1.1$ is rather narrow since the lightest stop mass varies fast for values of $X_t$ and $m_{\tilde t}\equiv m_{\tilde Q_3}=m_{\tilde u_3}$ that give an $r_\gamma$ enhancement.
This is reflected in a correspondingly rapid variation of $\delta r_G^{\tilde t}$ as a function of the soft supersymmetry breaking parameters.

The results of the fit are presented in the right panel of Figure~\ref{fig:stopsTheo}. The data currently prefer the narrow corridor described above. The $3\sigma$ contours instead lie both in the deeper minimum around the large $X_t$ region and in the high $m_{\tilde t}$ region where stop effects decouple. This second area corresponds to the shallower SM-like minimum of the fit shown in the right panel of Figure~\ref{fig:loops}.

Overall the results of the fit are in tension with the requirement of $10\%$ tuning, but point to a region directly accessible at the LHC. However a good fraction of the preferred masses falls where the sensitivity of current searches vanishes: $160\;\mathrm{GeV}\lesssim m_{\tilde t}\lesssim 220$ GeV~\cite{ATLASstops}, possibly motivating additional efforts.

\subsection{Charginos}\label{sec:charginos}
At least one chargino should be light $m_{\chi^\pm_1}\lesssim 300$~GeV in view of naturalness considerations (see Section \ref{sec:predict}), possibly giving an observable  deviation in the $h\to\gamma\gamma$ rate. However, the effects can be decoupled by raising $M_2$ to $\mathcal{O}(1\; \mathrm{TeV})$, without introducing more than a $20\%$ tuning~\cite{Papucci:2011wy}. Nonetheless, other than naturalness, there are further motivations to study chargino effects on Higgs couplings.

It possible that these deviations are the only observable effect of the MSSM at the LHC. Many models have the gauginos and higgsinos as the only light particles, with the possible addition of gluinos, as is the case of split supersymmetry~\cite{ArkaniHamed:2004fb, Giudice:2004tc} and models that address the moduli problem and dark matter \cite{Moroi:1999zb,Kaplan:2006vm,Acharya:2009zt}. The most studied scenarios typically decouple either gauginos or higgsinos~\cite{Hall:2012zp, Hall:2011jd}, which would make the effects on the Higgs couplings vanishingly small, but it is possible to obtain a viable dark matter candidate together with unification keeping both species light~\cite{Arvanitaki:2012ps}. Furthermore they are a reasonable proxy for a member of a $SU(2)_L$ multiplet providing a dark matter candidate. The mass splittings in the multiplets are expected to be of the order of a few hundreds of MeV~\cite{Cirelli:2005uq}, thus leaving open only monojet and monophoton searches at 14 TeV. These analyses may be the only viable path to direct exclusion/discovery also in the case of compressed SUSY spectra~\cite{Murayama:2012jh}. Their sensitivity is currently limited only to colored particles~\cite{Belanger:2012mk, Dreiner:2012gx, Drees:2012dd} and it was estimated that to probe the electroweak production of particles with masses up to 200 GeV, $300\;\mathrm{fb}^{-1}$ at 14 TeV will be necessary~\cite{Giudice:2010wb}. Therefore it is worth studying the chargino effects on the $h\gamma\gamma$ coupling in the MSSM, that were previously considered also in~\cite{Gunion:1988mf, Djouadi:1996pb, Diaz:2004qt, Arvanitaki:2012ps}.

\begin{figure}[t!]
\begin{center}
\includegraphics[width=0.45\textwidth]{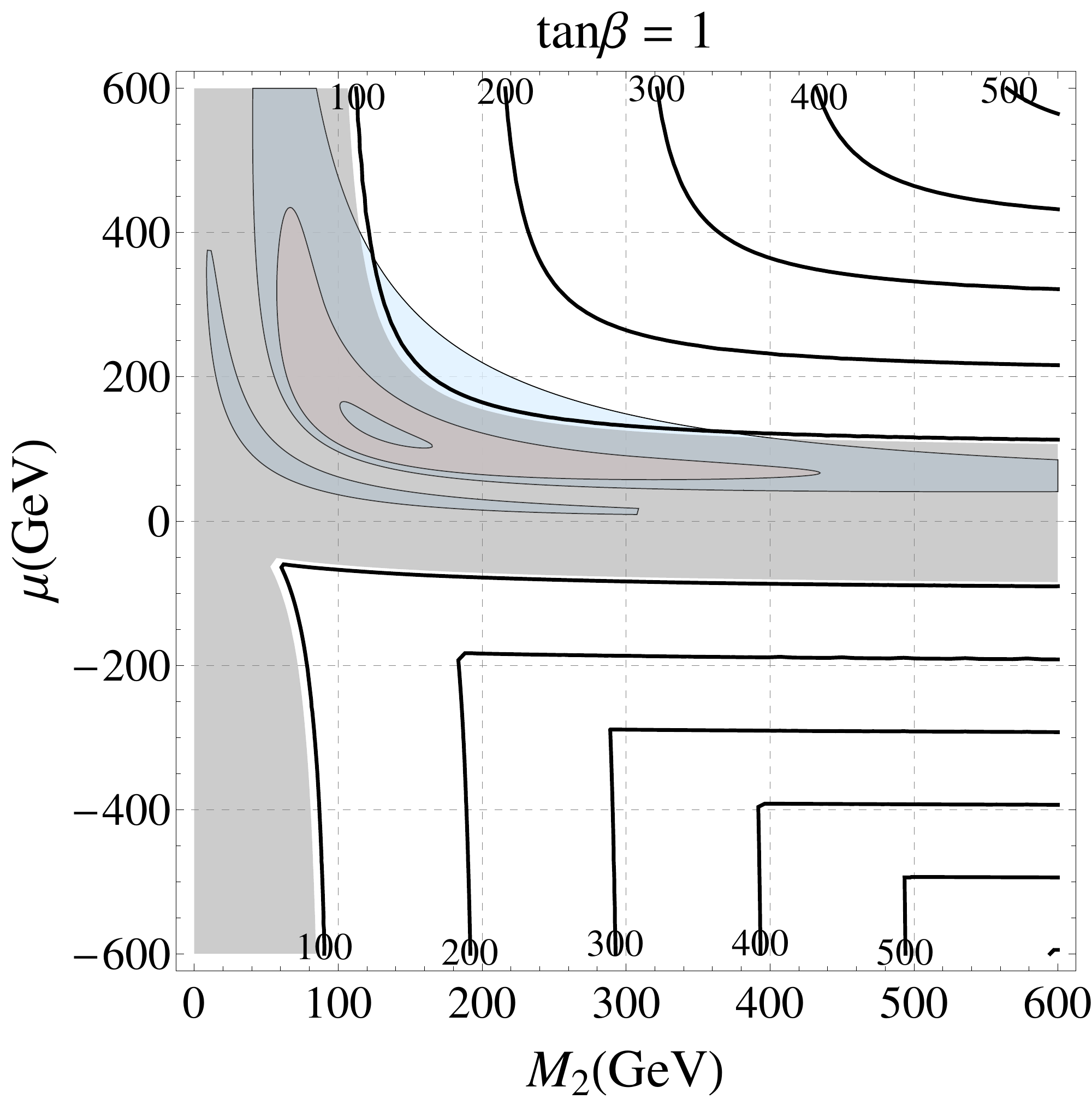}
\quad
\includegraphics[width=0.45\textwidth]{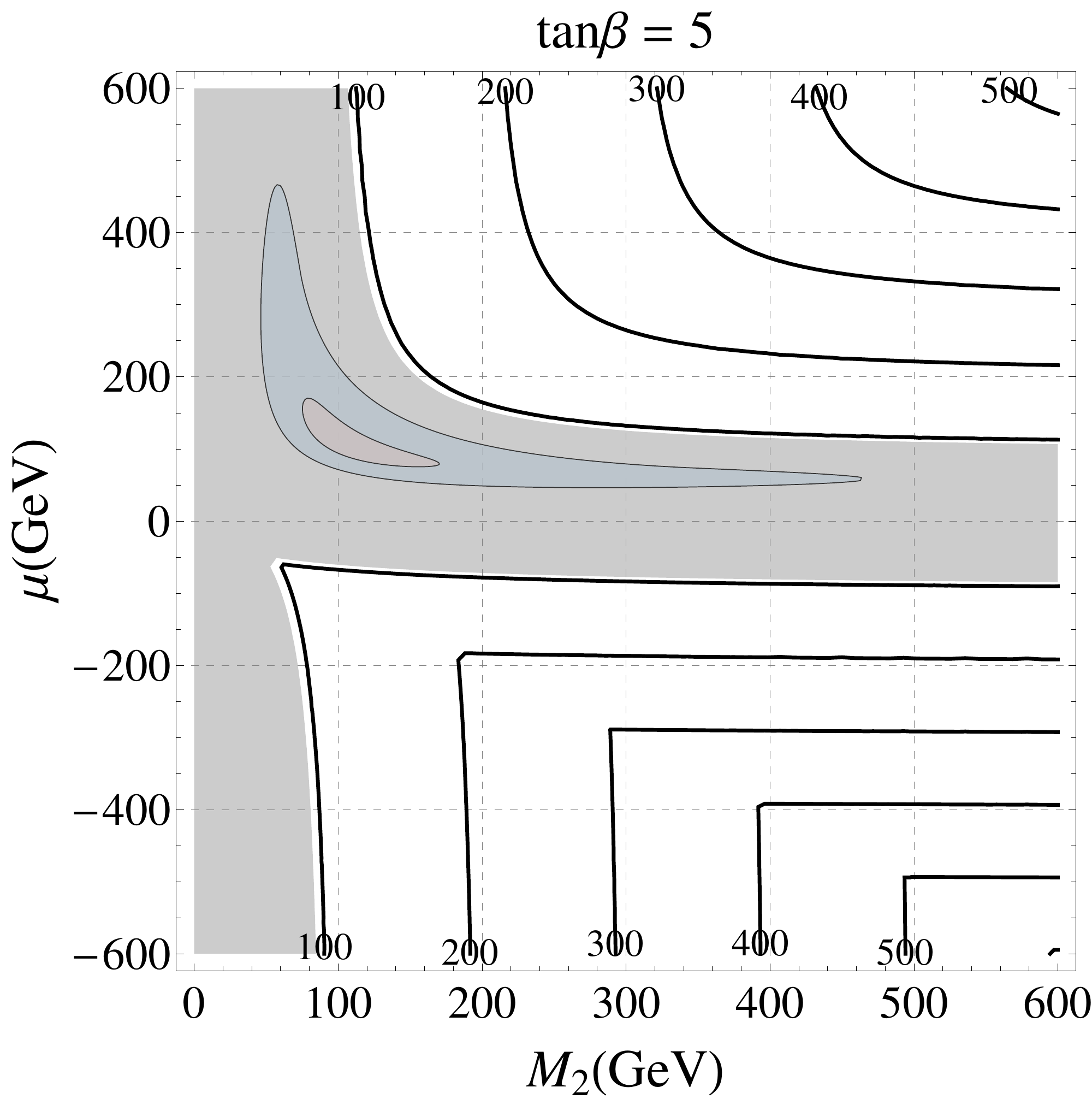}
\end{center}
\caption{Contours of the $\chi^2$ at $68\%$ and $95\%$ C.L. in the $(M_2, \mu)$ plane for $\tan\beta=1$ (left) and $\tan\beta=5$ (right). The black lines are contours of constant $m_{\chi_1^\pm}$ in GeV. In gray is shown the region excluded by LEP $m_{\chi_1^\pm}<94$~GeV.}
\label{fig:charginos}
\end{figure}%

In the MSSM, the chargino contribution is bounded by:  $-0.3 \leq \delta r^{\tilde{\chi}^\pm}_\gamma \leq 0.13$. It might be larger in the NMSSM~\cite{Choi:2012he} or in $D$-term extensions~\cite{Huo:2012tw}, but we do not consider this possibility here. Both limits come from direct chargino searches and can easily be derived by scanning $(M_2, \mu, \tan\beta)$ space with the LEP constraint $m_{\tilde{\chi}^\pm} > 94$ GeV \cite{Amsler:2008zzb}. The bounds are saturated when $\tan\beta =1$, restricting $\tan \beta \geq 2$ gives  $-0.2 \leq \delta r^{\tilde{\chi}^\pm}_\gamma \leq 0.1$~\cite{Blum:2012ii}.

It is thus clear that charginos alone can hardly explain the observed $h\to\gamma\gamma$ excess. The results of the fit in Figure~\ref{fig:charginos} roughly point to $\mu M_2 > \sin2\beta m_w^2 > 0$ where  the $h\gamma\gamma$ coupling is enhanced
\be
\delta r_\gamma^{\chi^\pm}\sim\frac{m_w^2 \sin 2\beta}{\mu M_2-m_w^2 \sin 2\beta}.~\label{eq:charg}
\ee
However the allowed region of the $(\mu, M_2)$ plane, even for $\tan\beta =1$, is at least $2\sigma$ away from the best fit to the data, while it disappears as $\tan\beta$ grows and the effect decouples. Note that the leading term in a $m_h^2/m_{\chi_1^\pm}^2$ expansion (equation~\ref{eq:charg}) ceases to be a good approximation in most of the region preferred by the fit, where one of the charginos can be extremely light $20-50$~GeV and already ruled out by LEP. As in the stops case, all the plots and the numbers that are quoted were obtained with the full MSSM one loop result~\cite{Djouadi:2005gj}.

\subsection{Five parameters fit}\label{sec:4D}
In the previous sections we  fit the data to simplified scenarios in which all the Higgs rates could be expressed in terms of one or two parameters. This gave an idea of the level of agreement between the data and some physically motivated corners of the parameter space of natural supersymmetry. In spite of the fact that only three channels per experiments have errors below the $50\%$ level (namely $\gamma\gamma$, $ZZ$ and $WW$ untagged) we still find it interesting to explore the more general case in which $r_\gamma,\; r_G,\; r_b, \; r_t$ and $r_V$ all play a role. This is a small modification of the four parameter natural MSSM, inspired by the possibility of adding a new singlet, and comes closer to approximating a fit with all couplings left to float. The only difference with respect to a four parameter fit resides in the fact that $r_t$ is virtually unconstrained and not artificially limited by its relation with $r_V$. These theory inspired exclusions can be applied also outside of the framework of natural supersymmetry and to facilitate possible attempts we show profiles of the $\chi^2$ for the five couplings in Appendix \ref{sec:chi2}.

\begin{table}[b!]
\caption{Confidence intervals for the five parameters that encode natural SUSY predictions for Higgs rates. All down-type couplings scale with $r_b$ and alll up-type ones with $r_t$. All other couplings not present in the table are fixed to their SM value.}
\begin{center}
\begin{tabular}{|c|c|}
\hline
&  95\% C.L. \textbf{(5D)} \\
\hline
$r_b$ & $0.96^{+0.64}_{-0.58}$  \\
\hline
$r_V$ &$0.96_{-0.33}^{+0.26}$\\
\hline
$ r_G$ &$0.89_{-0.30}^{+0.37}$\\
\hline
$ r_\gamma$ &$1.39^{+0.56}_{-0.50}$\\
\hline
$ r_t$ &$<2.07$\\
\hline
\end{tabular}
\end{center}
\label{tab:4dcoup}
\end{table}%

In Table~\ref{tab:4dcoup} we show the corresponding one dimensional 95\% C.L. intervals, obtained by treating the other parameters as nuisances. The errors on the single couplings vary from $30\%$ to $100\%$, which alone is not enough to lose all hope of constraining the parameter space of natural supersymmetry. However the impact of leaving all couplings to float is strong for charginos and the whole plane becomes accessible. Similarly for the case of stops, where the contribution in a large part of parameter space can be compensated by a shift in $r_t$.

\begin{figure}[t!]
\begin{center}
\includegraphics[width=0.45\textwidth]{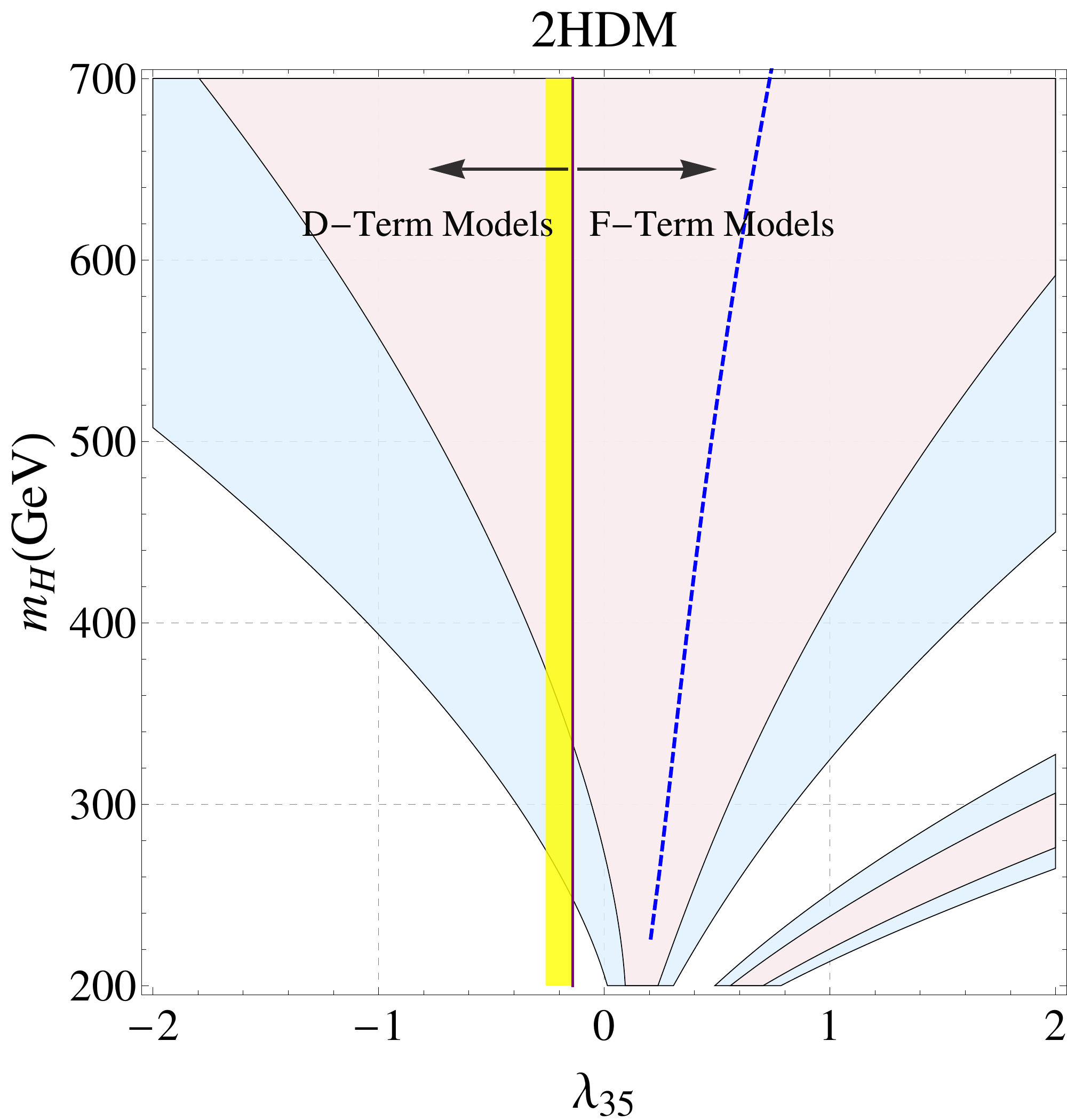}\quad
\end{center}
\caption{$\chi^2$ contours corresponding to the $68$\% and $95$\% confidence levels in the $(\lambda_{35}, m_H)$ plane. In the case of $r_b$ constrained by the five parameters fit ($r_G, r_\gamma, r_V, r_b, r_t$). The solid purple line corresponds to the tree-level value of $\lambda_{35}$ in the MSSM. The band in yellow covers the possible values of $\lambda_{35}$ in pure D-term models (defined in section~\ref{sec:res}). The blue dashed line runs through the best fit points.}
\label{fig:mssm2hdm4D}
\end{figure}

\begin{figure}[t!]
\begin{center}
\includegraphics[width=1\textwidth]{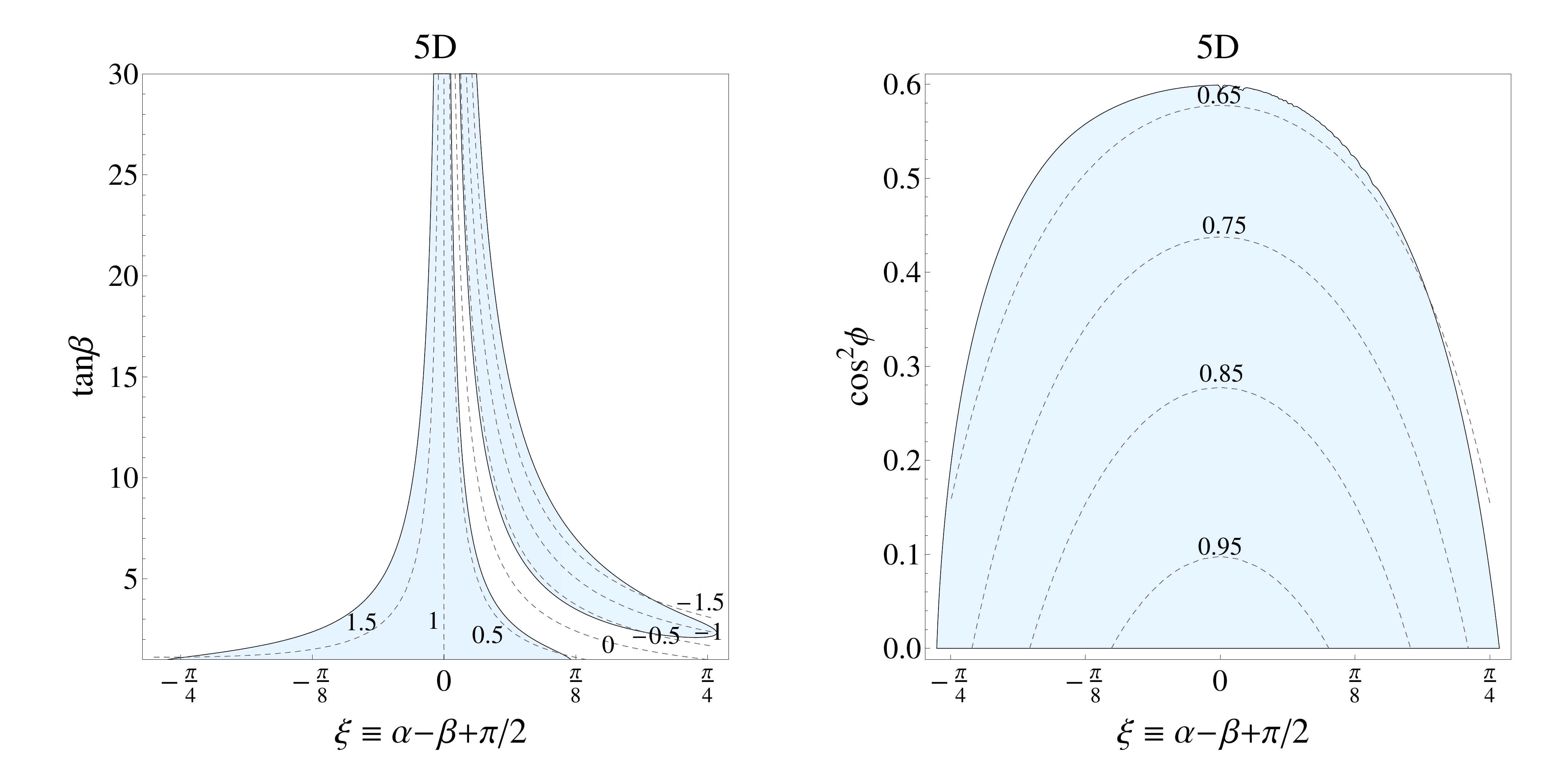}
\end{center}
\caption{{\bf Left:} The 95\% C.L. region in the $(\xi, \tan\beta)$ plane, obtained profiling the full five dimensional $\chi^2$. The dashed contours indicate $r_b$ for $\cos^2\phi=0$. {\bf Right:} The 95\% C.L. region in the $(\xi, \cos^2\phi)$ plane, obtained profiling the full five dimensional $\chi^2$. The dashed contours correspond to different values of $r_V$. }
\label{fig:xitb}
\end{figure}

On the other hand not all sensitivity is lost on tree-level couplings. For instance $\lambda_{35}=\lambda_{35}^{\mathrm{MSSM}}$ still gives a preferred region with $m_H \gtrsim 250$~GeV and $\lambda=1$ implies $m_H \gtrsim 320$~GeV, as it is shown in Figure~\ref{fig:mssm2hdm4D}. Similarly the tree-level statements about singlet mixing and the decoupling limit are meaningful. Repeating the exercise in Section~\ref{sec:mixing}, but profiling the full five dimensional $\chi^2$ we obtain $|\cos\phi|\lesssim 0.8$ at 95\% C.L. very close to the result of the purely tree-level fit.
This is an indication that both the tree-level fit and the five dimensional one are dominated by the lower bound on $r_V$ that for $\alpha \sim \beta -\pi/2$ \footnote{Note that this is the $\alpha$ defined in Equation~\ref{eq:matrix}.} becomes $r_V \approx \sin \phi$. The bound on $\cos\phi$ becomes much stronger away from the decoupling limit as can be seen in the right panel of Figure~\ref{fig:xitb}.

We can also extract information on the level of decoupling of the Higgs sector as a function of $\tan\beta$. In Figure~\ref{fig:xitb} the $95\%$ C.L. contour is plotted in the $(\xi, \tan\beta)$ plane. The key message is that large $\tan\beta$ is allowed only for values of $\xi$ close to decoupling, where corrections to $r_b\approx 1-\xi \tan\beta$ and $r_t \approx 1+\xi/ \tan\beta$ are small. This is reflected in the $(m_A, \tan\beta)$ exclusion discussed in section~\ref{sec:hmssm} that is even competitive with direct searches.
Many of these bounds apply to a vast class of complete theories and indicate that in the MSSM and many of its motivated extensions, tree-level effects in the Higgs sector are already strongly constrained, mainly by the measurements of the $h\to WW$ and $h\to ZZ$ rates that in our setting are always below their SM value unless $\xi=0$ and $\cos\phi=0$.

\section{Conclusion}
So far most of the effort has been devoted to simplified settings or complete models in which only one or two parameters enter Higgs rates. We set upon the more ambitious task of constraining a realistic theory using only naturalness to limit the number of parameters entering the game. Some of these ideas were treated already in~\cite{Blum:2012kn, Blum:2012ii}, but an application to the data was still missing.

We fit Higgs measurements, first treating independently the tree-level and loop effects, and then performing a five dimensional fit that parameterized complete natural theories such as the NMSSM, the MSSM and its $D$-term extensions. Considering a generic type-II 2HDM we found a bound on the heavy CP even Higgs mass $m_H \gtrsim 370$~GeV in theories with an approximate PQ symmetry. We also obtained a strong preference for the decoupling limit: $|\xi| \lesssim 0.1$, and in theories with an extra singlet a robust constraint on its mixing. When taking into account only loop effects we showed that the data prefer a very light stop ($\lesssim 200$~GeV) with large mixing, while charginos in the MSSM alone can not explain the current $h\to \gamma\gamma$ enhancement.

With the full five parameter fit we found that loop-level statements cease to be valid. However we were still able to draw interesting conclusions on tree-level mixings in the Higgs sector. We found that the mass bound in a type-II 2HDM with moderate to large $\tan\beta$ is relaxed to $m_H \gtrsim 250$~GeV and so is the upper bound on singlet mixing, $|\cos \phi|\lesssim 0.8$. However there is still a strong correlation between the size of $\tan\beta$ and the vicinity of the theory to the decoupling limit. This was translated into an exclusion in the $(m_A, \tan\beta)$ plane of the MSSM, that is competitive with direct searches. These bounds hold in a large class of complete theories, where loop corrections to the Yukawa couplings are small.

The comparison between the constraints discussed above and LHC searches indicates that there is still no tension between the two sets of measurements in natural supersymmetry and that direct searches are still the best chance of finding superpartners, barring extreme configurations with little visible and invisible energy in the event. The only exception are extra states in the MSSM Higgs sector, that at moderate $\tan\beta$ may be seen first as a deviation in Higgs couplings, rather than produced directly at the LHC.

\section*{Acknowledgements}
We thank Valerio Altini, Gideon Bella, Yonit Hochberg, Michele Papucci, Maurizio Pierini, Tomer Volansky, and Kathryn Zurek for useful discussions. The work
of EK is supported in part by a grant from the Israel Science Foundation.

\newpage

\appendix
\section{Results of the five dimensional fit}\label{sec:chi2}
\begin{figure}[!h]
\begin{center}
\includegraphics[width=0.45\textwidth]{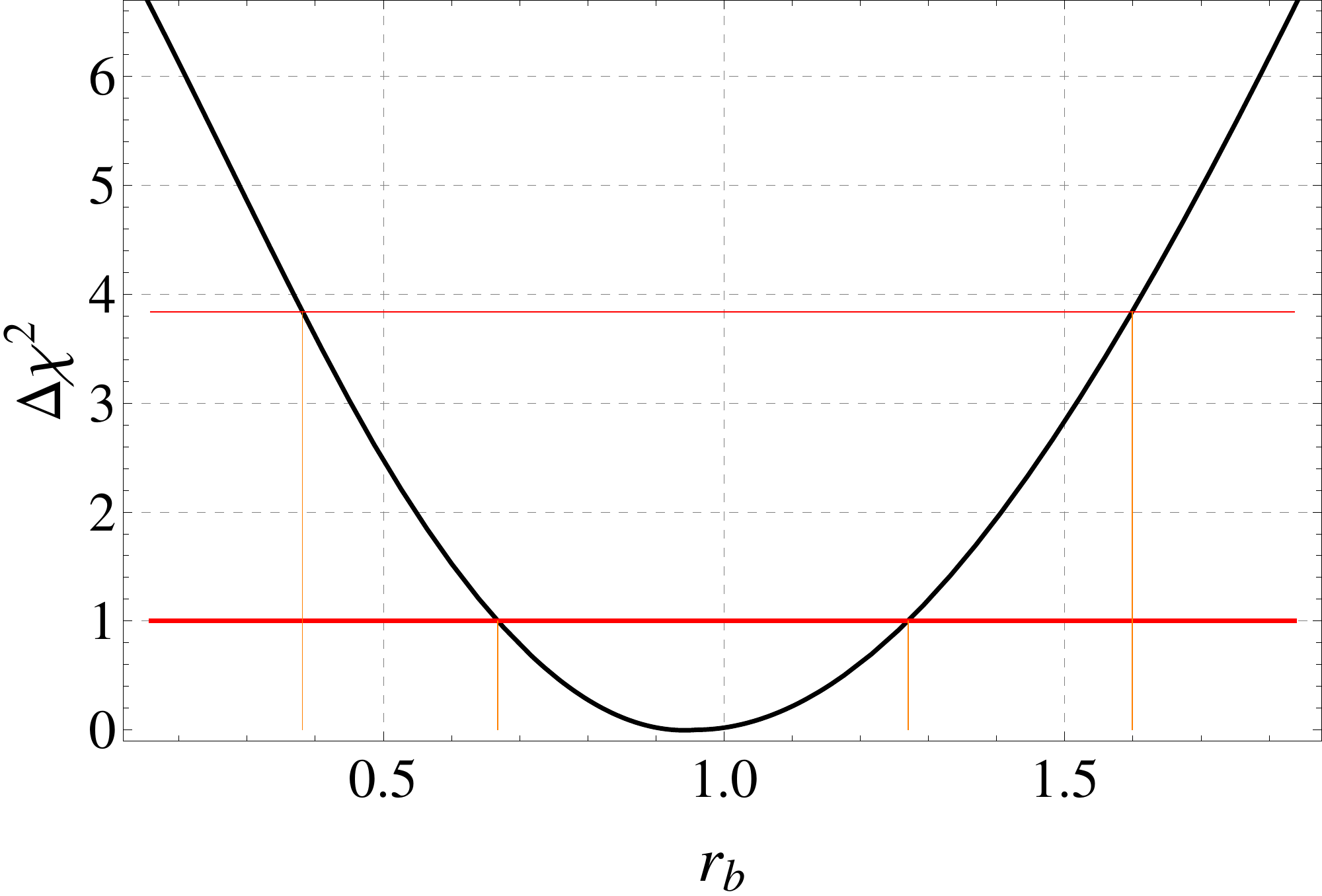}\quad
\includegraphics[width=0.45\textwidth]{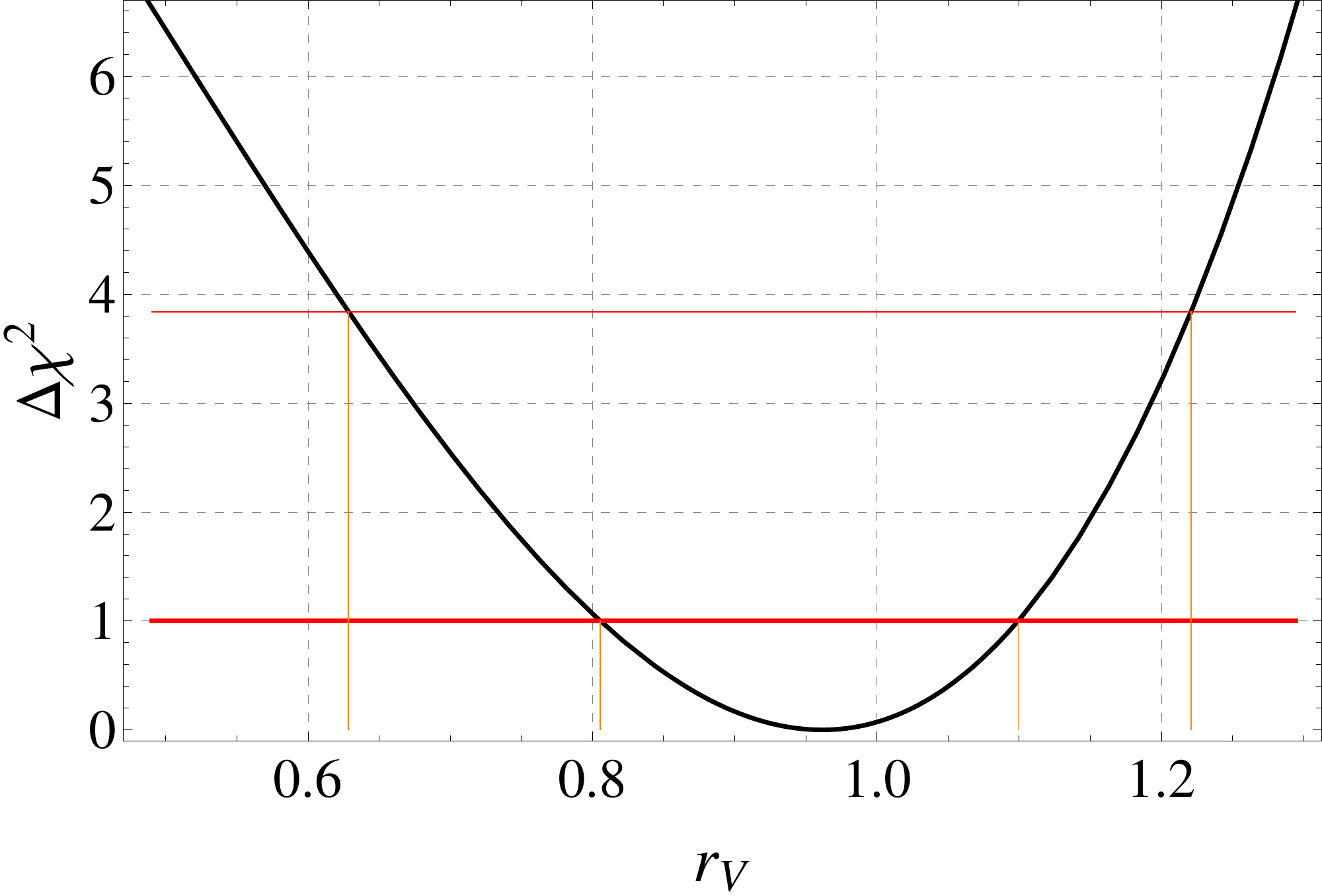}\quad
\includegraphics[width=0.45\textwidth]{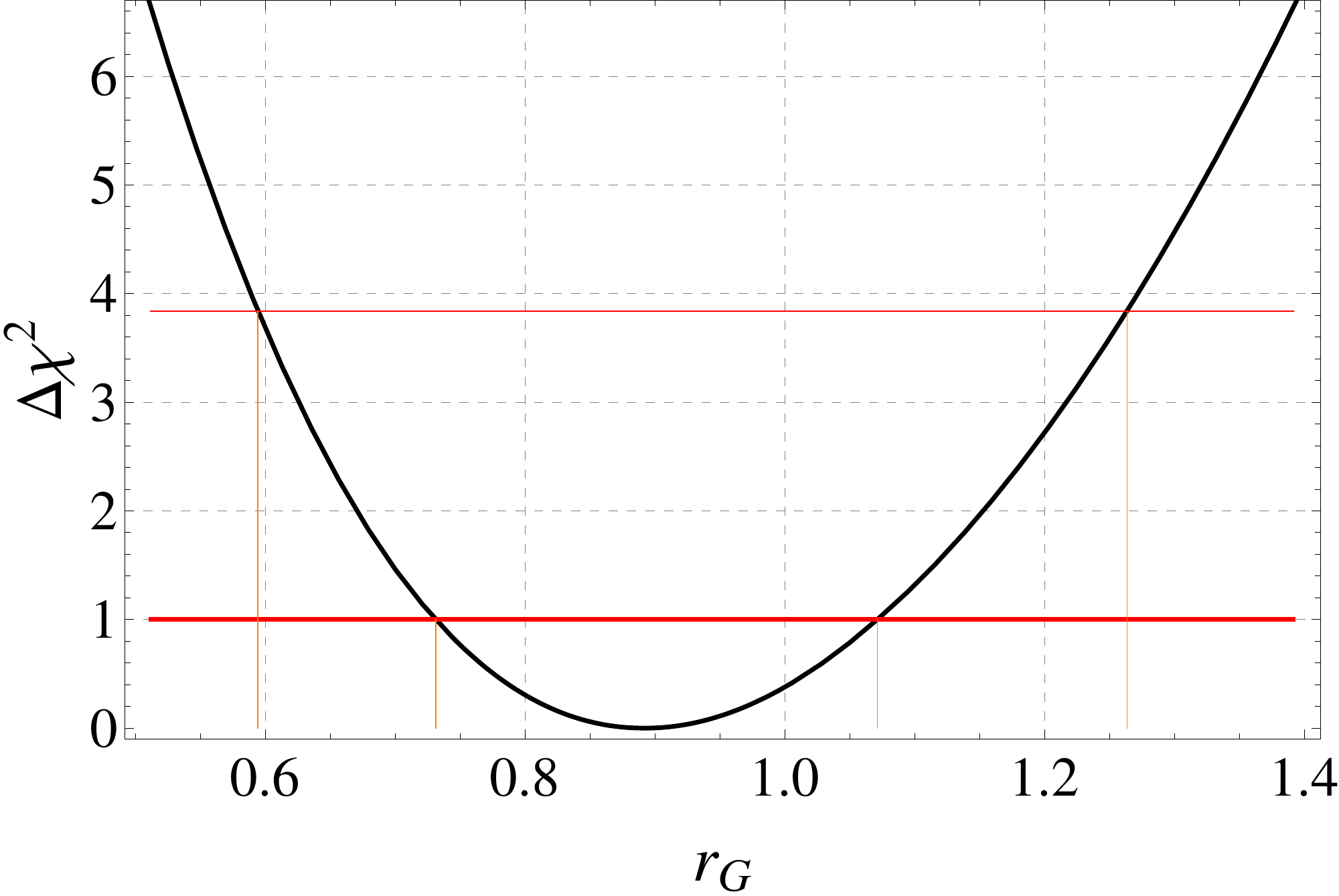}\quad
\includegraphics[width=0.45\textwidth]{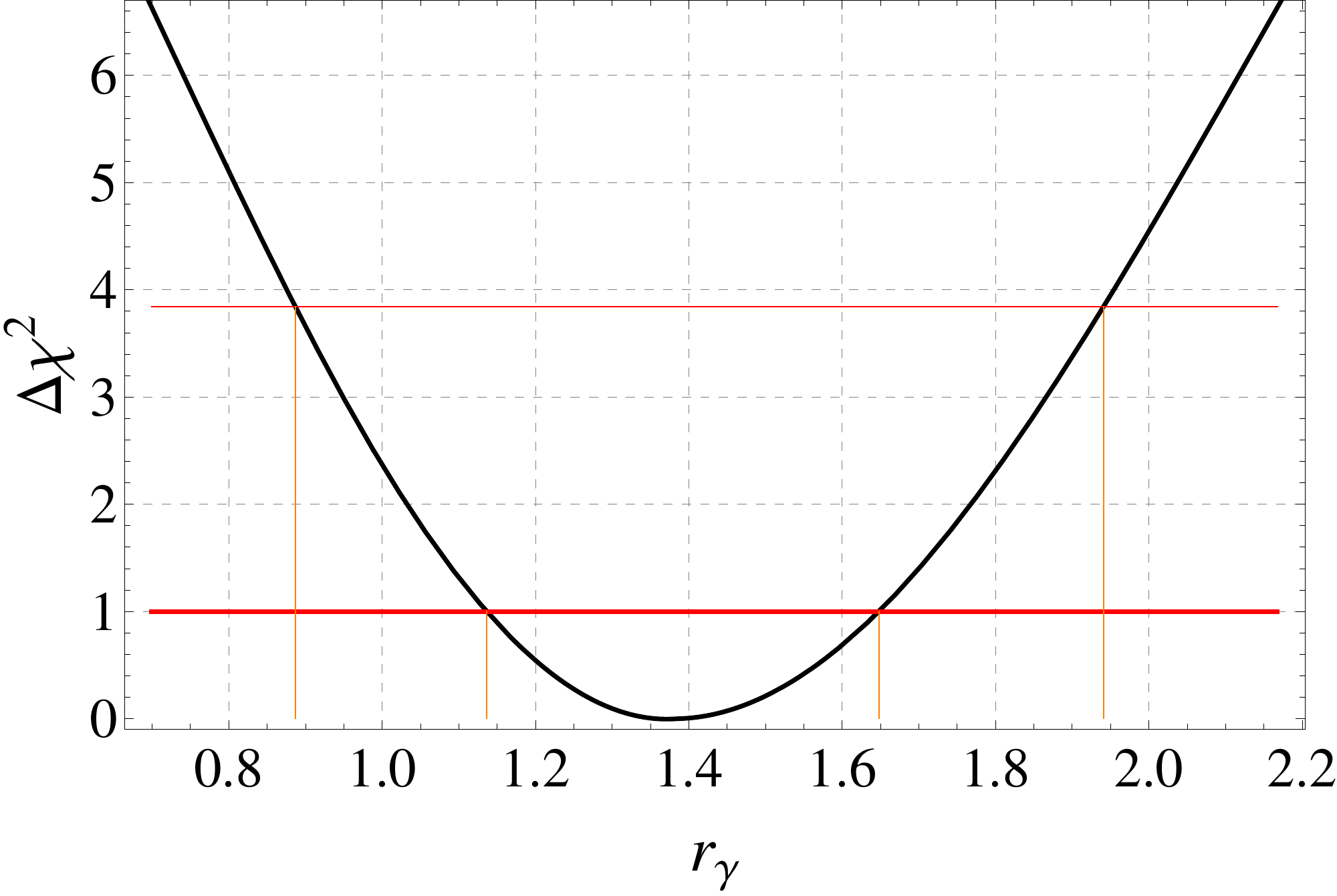}\quad
\includegraphics[width=0.45\textwidth]{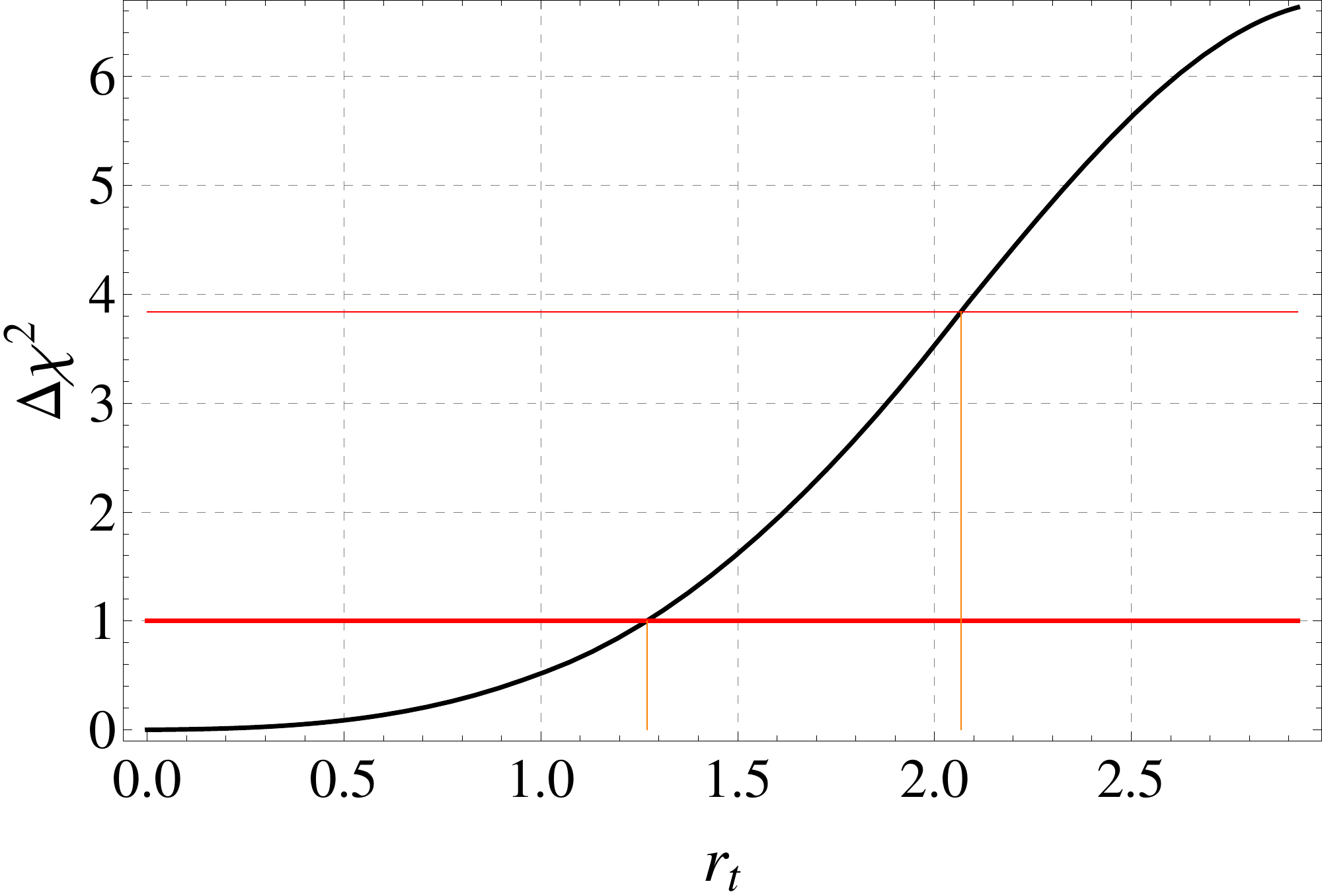}\quad
\end{center}
\caption{$\Delta \chi^2$ profiles for the five parameters fit ($r_G, r_\gamma, r_V, r_b, r_t$). Heights corresponding to the $68$\% and $95$\% confidence levels are marked in red.}
\label{fig:profiles}
\end{figure}

This appendix collects the one dimensional profiles of the $\chi^2$ obtained from the five dimensional fit described in section~\ref{sec:4D}. In Figure~\ref{fig:profiles} we show them with the $68\%$ and $95\%$ intervals marked in orange. The details of the profiling are discussed in section~\ref{sec:fit}. $r_t$, entering only at loop level in well measured rates, is essentially unconstrained due to the compensating effect of $r_G$ and $r_\gamma$.

\bibliography{ref}
\bibliographystyle{jhep}

\end{document}